\documentclass[twocolumn]{aastex631}

\usepackage{enumerate}
\usepackage{diagbox}
\usepackage{slashbox}
\usepackage{multirow}
\usepackage{amsmath}
\usepackage[utf8]{inputenc}
\usepackage{booktabs}
\let\oldequation\equation
\let\oldendequation\endequation

\usepackage{graphicx}
\usepackage{array}
\usepackage[figuresright]{rotating} 

\usepackage{graphicx}

\defcitealias{2022ApJ...925...31X}{Paper \uppercase\expandafter{\romannumeral1}}
\defcitealias{2022ApJ...939..104X}{Paper \uppercase\expandafter{\romannumeral3}}
\defcitealias{2023ApJ...944..200X}{Paper \uppercase\expandafter{\romannumeral4}}
\defcitealias{2024ApJ...969..129Z}{Paper \uppercase\expandafter{\romannumeral5}}

\renewenvironment{equation}{\linenomathNonumbers\oldequation}{\oldendequation\endlinenomath}

\shorttitle{PAC.\romannumeral7. Mass and Environment Quenching}
\shortauthors{Zheng et al.}

\begin{document}

\title{Photometric Objects Around Cosmic Webs (PAC). VII. Disentangling Mass and Environment Quenching with the Aid of Galaxy-halo Connection in Simulations}



\correspondingauthor{Y.P. Jing}
\email{ypjing@sjtu.edu.cn}

\author[0000-0001-6575-0142]{Yun Zheng}
\affiliation{State Key Laboratory of Dark Matter Physics, Tsung-Dao Lee Institute \& School of Physics and Astronomy, Shanghai Jiao Tong University, Shanghai 201210, People’s Republic of China}

\author[0000-0002-7697-3306]{Kun Xu}
\affiliation{Center for Particle Cosmology, Department of Physics and Astronomy,
University of Pennsylvania, Philadelphia, PA 19104, USA}
\affiliation{State Key Laboratory of Dark Matter Physics, Tsung-Dao Lee Institute \& School of Physics and Astronomy, Shanghai Jiao Tong University, Shanghai 201210, People’s Republic of China}
\affiliation{Institute for Computational Cosmology, Department of Physics, Durham University, South Road, Durham DH1 3LE, UK}

\author{Donghai Zhao}
\affil{College of Physics and Electronic Information Engineering, Guilin University of Technology, Guilin 541004, People’s Republic of China}
\affil{Key Laboratory of Low-dimensional Structural Physics and Application, Education Department of Guangxi Zhuang Autonomous Region, Guilin 541004, People’s Republic of China}
\affil{State Key Laboratory of Dark Matter Physics, Tsung-Dao Lee Institute \& School of Physics and Astronomy, Shanghai Jiao Tong University, Shanghai 201210, People’s Republic of China}

\author[0000-0002-4534-3125]{Y.P. Jing}
\affil{State Key Laboratory of Dark Matter Physics, Tsung-Dao Lee Institute \& School of Physics and Astronomy, Shanghai Jiao Tong University, Shanghai 201210, People’s Republic of China}

\author{Hongyu Gao}
\affil{State Key Laboratory of Dark Matter Physics, Tsung-Dao Lee Institute \& School of Physics and Astronomy, Shanghai Jiao Tong University, Shanghai 201210, People’s Republic of China}

\author[0000-0002-4574-4551]{Xiaolin Luo}
\affil{State Key Laboratory of Dark Matter Physics, Tsung-Dao Lee Institute \& School of Physics and Astronomy, Shanghai Jiao Tong University, Shanghai 201210, People’s Republic of China}

\author[0000-0002-1318-4828]{Ming Li}
\affil{National Astronomical Observatories, Chinese Academy of Sciences, Beijing, 100101, People’s Republic of China}

\begin{abstract}
Star formation quenching in galaxies plays a crucial role in galaxy evolution and is believed to be driven primarily by galaxy mass and environment. However, disentangling these effects is difficult because galaxy mass and environment are strongly correlated. In Paper V, we addressed this challenge using the Photometric Objects Around Cosmic Webs (PAC) method, which integrates spectroscopic and deep photometric surveys. This approach enabled us to measure the excess surface density \(\bar{n}_2w_{\rm{p}}(r_{\rm{p}})\) of blue and red galaxies surrounding massive central galaxies down to \(10^{9.0}M_{\odot}\). Despite this, fully separating mass and environmental effects remains challenging due to the difficulty in identifying a complete sample of low-mass central galaxies in redshift space. To overcome this, we derive the average quenched fraction of central galaxies, \(\bar{f}_{\rm{q}}^{\rm{cen}}(M_{*})\), by combining the reconstructed 3D quenched fraction distribution \(f^{\rm{sat}}_{\mathrm{q}}(r; M_{*,\rm{cen}}, M_{*,\rm{sat}})\), the stellar mass–halo mass relation in simulations, and the observed total quenched fraction, \(\bar{f}_{\rm{q}}^{\rm{all}}(M_{*})\). With \(\bar{f}_{\rm{q}}^{\rm{cen}}(M_{*})\), \(f^{\rm{sat}}_{\mathrm{q}}(r; M_{*,\rm{cen}}, M_{*,\rm{sat}})\) and the galaxy-halo connection, we assign a quenched probability to each (sub)halo and analyze galaxy quenching comprehensively. Our findings indicate that mass quenching dominates, with the mass-quenched fraction increasing from 0.3 to 0.87 across the stellar mass range \([10^{9.5}, 10^{11.0}]M_{\odot}\), while environmental quenching decreases from 0.17 to 0.03. Furthermore, more massive host halos are more effective at quenching their satellite galaxies, whereas satellite stellar mass has a minimal impact on environmental quenching.  Within the studied stellar mass range, \texttt{TNG300-1} exhibits lower mass-quenching efficiency but higher environmental-quenching compared to observations. 
\end{abstract}


\section{Introduction} \label{sec:intro}
Galaxies are now widely  classified into two primary populations: star-forming and passive. Galaxies belonging to the former category are typically young, actively producing new stars, and possessing blue colors and late-type morphologies. Those in the latter category  are typically old and red, have early-type morphologies, and do not exhibit signs of star formation \citep{2003ApJ...594..186B,2004ApJ...600..681B,2003MNRAS.341...54K,2004MNRAS.353..713K,2008A&A...483L..39C,2014ApJ...788...28V,2019MNRAS.483.5444D,2019MNRAS.488..847P}. In order to comprehend these galactic properties, a variety of physical mechanisms in galaxy formation and evolution should be considered. Among these, the quenching of star formation plays a critical role in shaping the properties of galaxies over cosmic time \citep{2003ApJ...594..186B,2004ApJ...600..681B,2004MNRAS.351.1151B,2008A&A...483L..39C,2013ApJ...777...18M,2019MNRAS.483.5444D,2019MNRAS.488..847P}. Therefore, a thorough examination of the galaxy quenching will eventually contribute significantly to our knowledge of the origin and evolution of galaxies.

The quenching of star formation in galaxies is intricately linked to both internal and external processes, which can be broadly categorized into mass quenching and environmental quenching mechanisms \citep{2010ApJ...721..193P,2010MNRAS.402.1942C,2011MNRAS.411..675S,2012ApJ...757....4P,2013ApJ...777...18M,2016ApJ...825..113D,2016MNRAS.457.4360Z,2017MNRAS.464..121S,2019MNRAS.483.2851S,2020ApJ...889..156C,2020ApJ...890....7C,2022A&A...668A..69E,2023arXiv231210222T}. Mass quenching, also known as internal quenching, primarily involves processes that are dependent on the galaxy's stellar mass, such as gas outflows due to supernova explosions and stellar winds \citep{1974MNRAS.169..229L,1986ApJ...303...39D,2008MNRAS.387.1431D} and active galactic nucleus (AGN) feedback from the supermassive black hole \citep{2006MNRAS.365...11C,2007ApJ...660L..11N,2012ARA&A..50..455F,2013ApJ...776...63F,2014A&A...562A..21C,2018MNRAS.476...12B}. On the other hand, environmental quenching is driven by interactions between galaxies and their surroundings, such as ram pressure stripping  \citep{1972ApJ...176....1G,1999MNRAS.304..465M,2017MNRAS.466.1275B,2017ApJ...844...48P,2018ApJ...857...71B,2019ApJ...873...52O,2021PASA...38...35C}, strangulation or starvation  \citep{1980ApJ...237..692L,1999MNRAS.304..465M,2011ApJ...732...17N,2015Natur.521..192P}, and harassment  \citep{1981ApJ...243...32F,1996Natur.379..613M}.

The relative importance of mass quenching and environmental quenching in the cessation of star formation in galaxies remains a topic of active debate. While mass quenching is widely accepted as the dominant process in massive central galaxies, the role of environmental quenching is less clear. Numerous observational studies have demonstrated that galaxies are more likely to be quenched in denser environments  \citep{2000ApJ...540..113B,2007ApJ...664..791B,2014ApJ...789..164T,2017MNRAS.464..121S,2020ApJ...889..156C}. However, other studies have reported little to no dependence on environmental proxies such as halo mass and cluster-centric distance  \citep{2012ApJ...746..188M,2016ApJ...825..113D,2018MNRAS.475..523L}. 

To comprehensively understand the effects of mass and environmental quenching, two major challenges must be addressed. First, accurately measuring the environment of galaxies is very important, as massive galaxies are found always in high-density regions so the galactic mass and the environment are strongly coupled.  Furthermore, the redshift space distortion makes the measurement of the over-density in redshift galaxy surveys very difficult. Second, accurately measuring the properties of low-mass galaxies is crucial, as environmental quenching predominantly impacts this population. While some recent studies have demonstrated progress in separating mass-driven and environment-driven quenching processes \citep{2010ApJ...721..193P,2012ApJ...746..188M,2014MNRAS.438..717K,2016ApJ...825..113D,2018MNRAS.475..523L,2022A&A...666A.141M}, and shown that these processes are largely separable \citep{2010ApJ...721..193P,2012ApJ...744...88Q,2014MNRAS.438..717K,2018A&A...618A.140V}, accurately quantifying  the effects of the mass quenching and environment quenching remains challenging. This difficulty stems from the complexities of identifying a complete sample of central galaxies and of accurately defining their environment in redshift space.

To tackle these challenges, \cite{2024ApJ...969..129Z} (hereafter \citetalias{2024ApJ...969..129Z}) employed the Photometric Objects Around Cosmic Webs (PAC) method \citep{2022ApJ...925...31X}. This approach integrates data from cosmological spectroscopic and photometric surveys, leveraging the depth of photometric surveys to extend measurements of galaxy properties and distributions to much lower mass ranges. \citetalias{2024ApJ...969..129Z} estimated the excess surface distribution $\bar{n}_2w_{\mathrm{p}}(r_{\mathrm{p}})$ of photometric galaxies in different stellar mass bins ($10^{9.5}M_{\odot}<M_*<10^{11.0}M_{\odot}$) and colors around spectroscopic massive central galaxies ($10^{10.9}M_{\odot}<M_*<10^{11.7}M_{\odot}$) at $z_s<0.2$, using Slogan Digital Sky Survey \citep[SDSS;][]{2000AJ....120.1579Y} spectroscopic and photometric samples. The measurements do not suffer from the redshift distortion, thus providing an accurate quantification of environments in real space. \citetalias{2024ApJ...969..129Z} did not extend the measurements to lower mass centrals due to high contamination issues mentioned above. \citetalias{2024ApJ...969..129Z} also provided the measurements at higher redshift using SDSS LOWZ ($0.3<z_s<0.5$) and CMASS ($0.5<z_s<0.7$) spectroscopic samples \citep{2015ApJS..219...12A,2016MNRAS.455.1553R} and Hyper Suprime-Cam Subaru Strategic Program (HSC-SSP) photometric catalogs \citep{2019PASJ...71..114A}. Based on the $\bar{n}_2w_{\mathrm{p}}(r_{\mathrm{p}})$ measurements for different colors, \citetalias{2024ApJ...969..129Z} calculated projected quenched fraction and projected quenched fraction excess (QFE) of companion galaxies around central galaxies. \citetalias{2024ApJ...969..129Z} concluded that the high-density host halo environment influences the star formation of companion galaxies up to a scale of approximately 3 time the viral radius $r_{\rm{vir}}$. \citetalias{2024ApJ...969..129Z} also studied dependence of QFE on central/companion mass and provide a fitting formula to describe all these dependencies.

However, the projected QFE in \citetalias{2024ApJ...969..129Z} was calculated by subtracting the average quenched fraction, \(\bar{f_{\rm{q}}}\), measured around \(r_{\rm{p}} = 3r_{\rm{vir}}\), which was assumed to represent the effects of mass alone. This assumption is not entirely accurate, as \(\bar{f_{\rm{q}}}\) at \(r_{\rm{p}} = 3r_{\rm{vir}}\) includes a combination of mass effects and the influence of the average environment at that scale. Consequently, although \citetalias{2024ApJ...969..129Z} achieved significant progress in the measurements, the effects of mass and environment on the quenching of companion galaxies have not been fully disentangled. In this paper, to further address this remaining issue, we first recover the 3D quenched fraction distribution, \(f^{\rm{com}}_{\rm{q}}(r;M_{*,\rm{cen}},M_{*,\rm{com}})\), of companion galaxies from the measured \(\bar{n}_2w_{\mathrm{p}}(r_{\mathrm{p}})\) in different stellar mass bins, assuming power-law galaxy distributions. Using an N-body simulation and the precise stellar mass–halo mass relation (SHMR) from \cite{2023ApJ...944..200X} (hereafter \citetalias{2023ApJ...944..200X}), we assign colors to satellite galaxies (within \(r_{\rm{vir}}\)) in the simulation based on \(f^{\rm{com}}_{\rm{q}}(r;M_{*,\rm{cen}},M_{*,\rm{com}})\) within each halo. Next, we calculate the quenched fraction of central galaxies in each stellar mass bin by combining the mean quenched fraction of satellites, $\bar{f}^{\rm{sat}}_{\rm{q}}$, derived from the simulation, the central and satellite galaxy numbers from the simulation, and the mean quenched fraction of all galaxies, $\bar{f}^{\rm{all}}_{\rm{q}}$, from the observation data. Finally, we disentangle and quantify the effects of mass and environment on galaxy quenching by comparing the quenching fractions of central and satellite galaxies.

The structure of this paper is as follows. In Section \ref{sec:Data and Method}, we give an overview of the observational data and the simulation analyzed in this work. Section \ref{sec:model} outlines the development of our model to disentangle the effects of mass and environment. In Section \ref{sec:quenching}, we explore how the quenched fraction varies with the environment and stellar mass, after which we draw our conclusions in Section \ref{sec:Coclusion}. Throughout the paper we adopt the Planck 2018 $\Lambda$CDM  model \citep{2020A&A...641A...6P} with cosmological parameters as $\Omega_{\mathrm{m},0} = 0.3111$, $\Omega_{\Lambda,0} = 0.6889$ and $H_0 = 67.66 \,\mathrm{km\,s^{-1}\,Mpc^{-1}}$ .

\begin{table*}
\centering
\caption{The marginalized posterior PDFs of the parameters. $M_0$ is in units of $h^{-1}M_{\odot}$ and $k$ is in units of $M_{\odot}$.}\label{DP model}
\begin{tabular}{cccccc}\\
\hline \hline  galaxies  & ${\mathrm{log}_{10}}(M_0)$ & $\alpha$ & $\beta$ & ${\mathrm{log}}_{10}(k)$ & $\sigma$ \\
\hline 
\text{central} & $11.716_{-0.048}^{+0.050}$ & $0.356_{-0.021}^{+0.020}$ & $2.422_{-0.106}^{+0.201}$ & $10.201_{-0.047}^{+0.045}$ & $0.226_{-0.013}^{+0.013}$ \\

\text{satellite} & $11.943_{-0.066}^{+0.074}$ & $-0.012_{-0.236}^{+0.148}$ & $1.711_{-0.033}^{+0.034}$ & $10.492_{-0.070}^{+0.086}$ & $0.307_{-0.032}^{+0.022}$ \\

\hline
\end{tabular}
\end{table*}

\begin{figure}[htb!]
\includegraphics[width=0.5\textwidth]{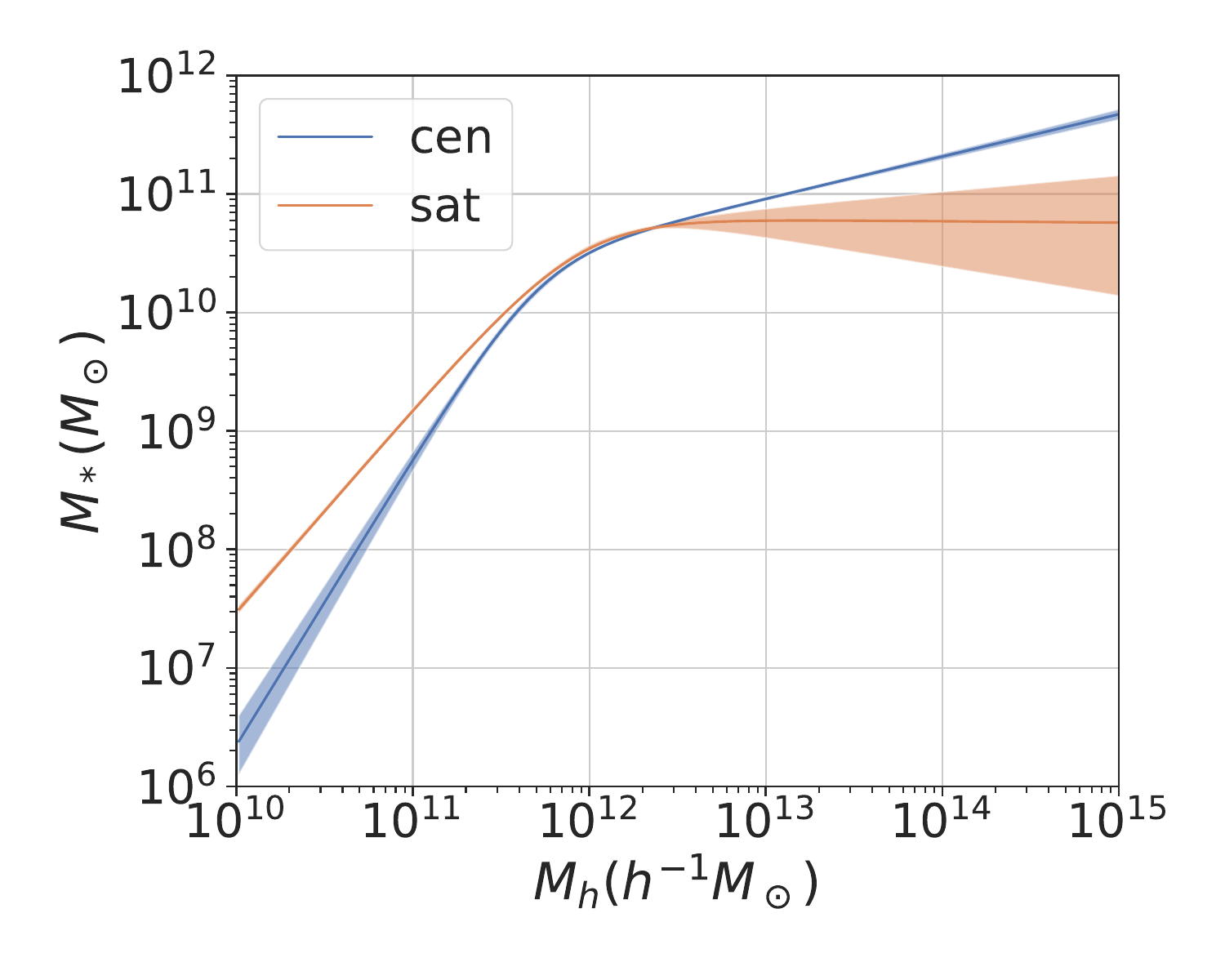}
\caption{The mean SHMRs (lines) and their $1\sigma$ errors (shadows) for central galaxies and satellite galaxies from SHAM. The orange line is for satellite galaxies and the blue line is for central galaxies.}
\label{SHMR}
\end{figure}

\begin{figure}[htb!]
\includegraphics[width=0.5\textwidth]{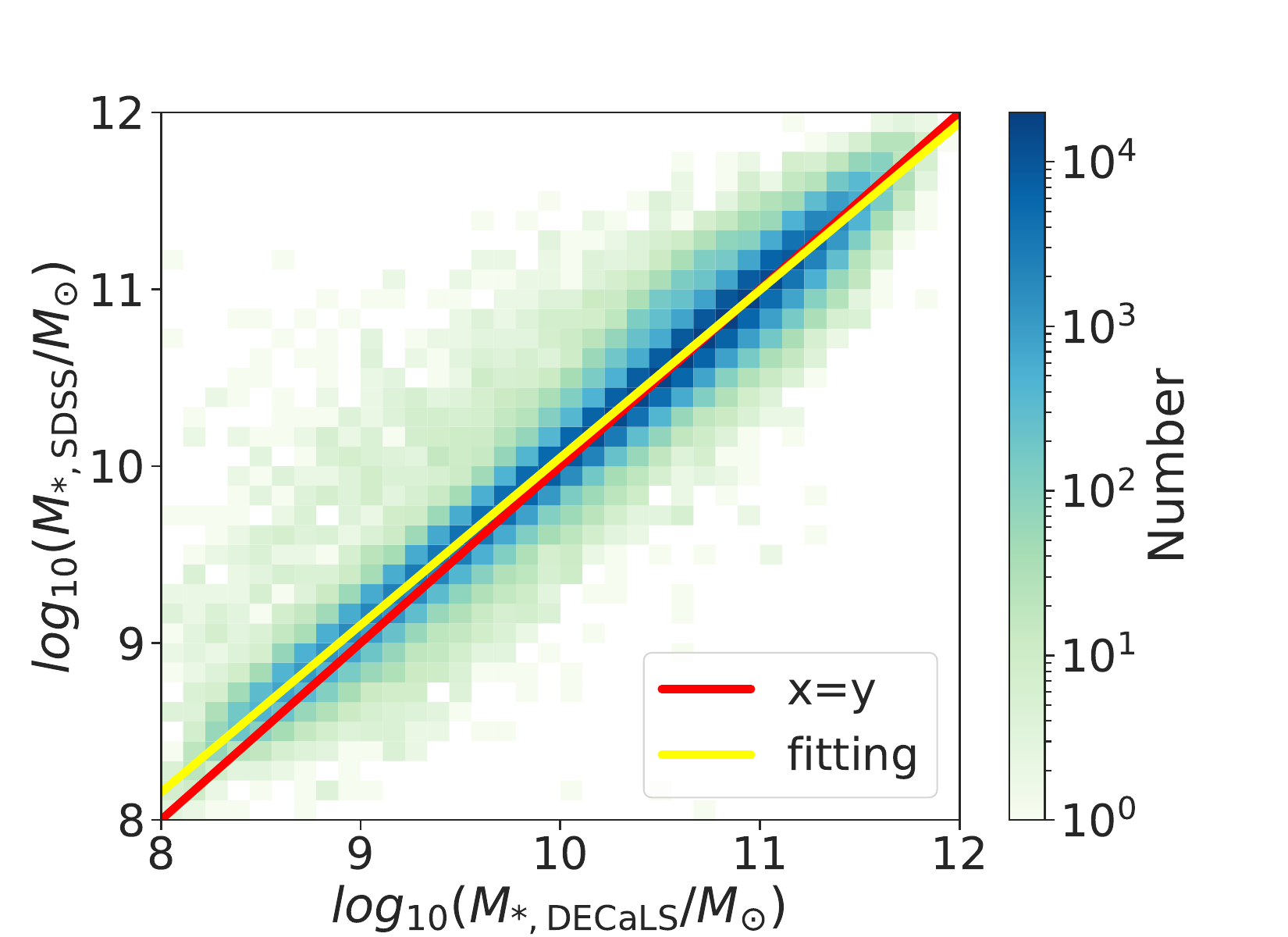}
\caption{The comparison of the stellar masses of the identical galaxies obtained by running the SED fitting in two distinct photometry, as depicted by the labels. The color represents the number of galaxies. The yellow line is the fitted result to illustrate the correlation between the two datasets. The red line shows $x=y$ for reference.}
\label{SDSS_decals}
\end{figure}

\section{Observational and simulation data} \label{sec:Data and Method}
In this section, we present the observational and simulation data utilized to construct the galaxy quenching model. While largely based on \citetalias{2024ApJ...969..129Z}, these data have undergone more precise calibration to achieve improved consistency between observations and simulations.

\subsection{Simulation and SHMR}
We use the {\tt{Jiutian}} simulation and precise SHMR to build mock catalogs for galaxies with different stellar masses. 

The {\tt{Jiutian}} suite consists of a sequence of N-body simulations created to satisfy the scientific needs of the Chinese Space Station Telescope optical surveys \citep{2019ApJ...883..203G}. We employ one of the high-resolution main runs based on the Planck 2018 cosmology  \citep{2020A&A...641A...6P}, as mentioned at the end of the Introduction. This simulation contains $6144^3$ dark matter particles within a periodic box of $1000 h^{-1}\mathrm{Mpc}$, performed with the GADGET-3 code \citep{2001NewA....6...79S}. The Friends-of-Friends technique is used to identify dark matter halos, and {\tt{HBT+}} is used to track subhalos  \citep{2012MNRAS.427.2437H,2018MNRAS.474..604H}. We use the snapshot at $z_s = 0.102$ to compare with the SDSS observational data at $z_s<0.2$.

We adopt the measurements and a similar methodology from \citetalias{2023ApJ...944..200X} to derive the SHMR. \citetalias{2023ApJ...944..200X} measured 95 $\bar{n}_2 w_{\mathrm{p}}(r_{\mathrm{p}})$ across various stellar mass bins, reaching down to $10^{8.0}M_{\odot}$, using the Dark Energy Camera Legacy Survey (DECaLS) photometric catalog and the SDSS Main spectroscopic sample at $z_s < 0.2$. Here, $\bar{n}_2$ represents the mean number density of photometric galaxies, and $w_{\mathrm{p}}$ is the projected cross-correlation function of spectroscopic and photometric galaxies. By modeling these $\bar{n}_2 w_{\mathrm{p}}(r_{\mathrm{p}})$ measurements using subhalo abundance matching (SHAM) with a parameterized SHMR, \citetalias{2023ApJ...944..200X} achieved $1\%$ precision in constraining the parameters. In this work, we follow their methodology and use the same measurements but introduce separate SHMRs for central and satellite galaxies, unlike the unified SHMR approach in \citetalias{2023ApJ...944..200X}. We find that this refined model provides a better fit to the measurements and offers a more physically reasonable interpretation (K. Xu et al. 2025, in prep.).

we use the double power law form (the DP model in \citetalias{2023ApJ...944..200X}):

\begin{equation}
M_*=\left[\frac{2k}{\left(\frac{M_{\rm acc}}{M_0}\right)^{-\alpha}+\left(\frac{M_{\rm acc}}{M_0}\right)^{-\beta}}\right].
\end{equation}

Here $M_{\mathrm{acc}}$ refers to the viral mass $M_{\mathrm{vir}}$ of the halo at the specific time when the galaxy was last to be the central dominant object. $M_*$ represents the stellar mass. A Gaussian function with width $\sigma$ is adopted to describe the dispersion in $\log(M_*)$ at a given $M_{\mathrm{acc}}$. The slopes of the SHMR at the high and low mass ends are represented by $\alpha$ and $\beta$, respectively. As we mentioned before, we use two different SHMRs for the central and satellite galaxies separately to model the $\bar{n}_2 w_{\mathrm{p}}(r_{\mathrm{p}})$ measurements from \citetalias{2023ApJ...944..200X}. The constrained parameters for the DP model for the central and satellite galaxies are given in Table \ref{DP model} and the corresponding SHMRs are shown in Figure \ref{SHMR}.  The fits are shown in Figure \ref{fig:PAC_fit}, demonstrating excellent agreement with the 95 cross-correlations, with a reduced chi-square value of \(\chi^2/{\rm{d.o.f.}} \sim 0.8\), confirming the accuracy and reliability of our model. 

However, in \citetalias{2023ApJ...944..200X}, the stellar mass is calculated using the SED code CIGALE \citep{2019A&A...622A.103B} based on DECaLS photometry with $g$, $r$ and $z$ bands, while the stellar mass used in \citetalias{2024ApJ...969..129Z} to study galaxy quenching is based on SDSS photometry with five bands $ugriz$. Although the SED fitting code and models are the same, we still find some small differences. In Figure \ref{SDSS_decals}, we compare the stellar masses of identical galaxies derived from the two different photometry, and find a linear relation can well describe their differences: 
\begin{equation}
\begin{aligned}
    &{\mathrm{log}}_{10}(M_{\mathrm{*,SDSS}}/M_{\odot}) \\
    &= 0.946{\mathrm{log}}_{10}(M_{\mathrm{*,DECLaS}}/M_{\odot}) +0.584.
\end{aligned}\label{eq:stellar mass}
\end{equation}
After assigning the DECaLS stellar masses to halos and subhalos in the {\tt{Jiutian}} simulation, we further calibrated them to SDSS stellar masses using Equation \ref{eq:stellar mass}. This calibration ensures a more consistent comparison with observational data from \citetalias{2024ApJ...969..129Z}.

\begin{figure*}[!htp]
\begin{center}
\includegraphics[width=0.8\textwidth]{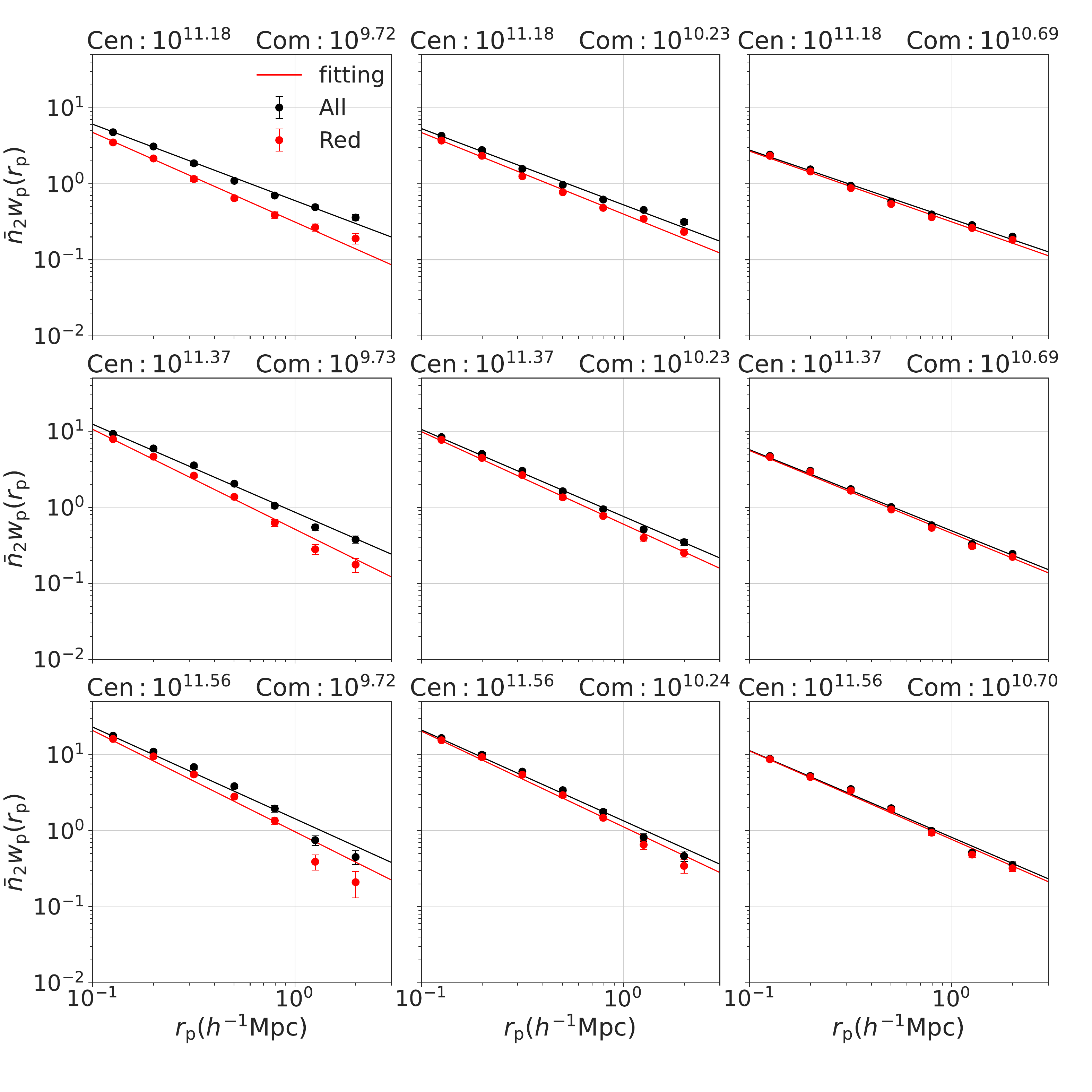}
\end{center}
\caption{The measurements of \(\bar{n}_2w_{\mathrm{p}}(r_{\mathrm{p}})\) are shown for all galaxies (in black) and red subsamples (in red). Dots with error bars represent observations, while lines indicate the fitting results. The subplots, arranged from left to right, correspond to companion mass bins with edges \([10^{9.5}, 10^{10.0}]\), \([10^{10.0}, 10^{10.5}]\), and \([10^{10.5}, 10^{11.0}] M_{\odot}\). From top to bottom, the subplots correspond to central galaxy mass bins with edges \([10^{11.1}, 10^{11.3}]\), \([10^{11.3}, 10^{11.5}]\), \([10^{11.5}, 10^{11.7}]\), and beyond. Each subplot title specifies the central and companion stellar mass bin centers.}
\label{fit_nwp}
\end{figure*}

\subsection{Measuring $\bar{n}_2w_{\mathrm{p}}(r_{\mathrm{p}})$ across different colors and stellar masses}\label{SDSS_sample}
We use the same data and methodology as \citetalias{2024ApJ...969..129Z} to obtain the projected density distribution \(\bar{n}_2 w_{\mathrm{p}}(r_{\mathrm{p}})\) of companion galaxies with different colors and stellar masses around massive central galaxies with varying stellar masses. However, we apply different criteria to select central galaxies to ensure greater consistency with the definition of central galaxies in the simulation.

In this work, we use the SDSS DR7 Main spectroscopic sample \citep{2009ApJS..182..543A} and DR13 {\tt{datasweep}} photometric sample \citep{2017ApJS..233...25A}, focusing on galaxies with redshifts \(z_s < 0.2\). In \citetalias{2024ApJ...969..129Z}, we selected central galaxies in the Main sample as those that do not have more massive neighbors within a distance of $30r^{\rm nb}_{\mathrm{vir}}$ along the line-of-sight (LOS) and within $3r^{\rm nb}_{\mathrm{vir}}$ perpendicular to the LOS, based on our finding that the environmental impact of a halo extends up to approximately $3r_{\mathrm{vir}}$. Here, $r^{\rm nb}_{\mathrm{vir}}$ refers to the virial radius of the more massive galaxies for each comparison. In this paper, to be consistent with the definition used in the simulation, we adjust the selection criteria by applying $r^{\rm nb}_{\mathrm{vir}}$ perpendicular to the LOS as the threshold, while keeping the selection distance along the LOS the same. This adjustment is made because centrals in the simulation are typically defined within each halo, making $r_{\mathrm{vir}}$ a more appropriate choice for comparison. We calculate \(r_{\mathrm{vir}}\) for each central stellar mass bin we are interested in based on the SHMR and the {\tt{Jiutian}} simulation. The results are \(0.47 \pm 0.194\,h^{-1}{\rm{Mpc}}\), \(0.631 \pm 0.263\,h^{-1}{\rm{Mpc}}\), and \(0.835 \pm 0.329\,h^{-1}{\rm{Mpc}}\) for the central stellar mass bins \([10^{11.1} M_{\odot}, 10^{11.3} M_{\odot}]\), \([10^{11.3} M_{\odot}, 10^{11.5} M_{\odot}]\), and \([10^{11.5} M_{\odot}, 10^{11.7} M_{\odot}]\), respectively.

We use the same color cut as \citetalias{2024ApJ...969..129Z} to define blue and red galaxies based on the rest-frame $u-r$ color:
\begin{equation}
u - r  = 0.11 \log M_{\odot} + 0.895. \label{colorcut_eq}
\end{equation}
Then, we calculate \(\bar{n}_2 w_{\mathrm{p}}(r_{\mathrm{p}})\) for the red and the entire companion galaxy sample in four stellar mass bins within $[10^{9.5}M_{\odot},10^{11.0}M_{\odot}]$ around massive central galaxies in four stellar mass bins within $[10^{11.1}M_{\odot},10^{11.7}M_{\odot}]$ . We adopt the jackknife resampling technique to assess the statistical error. The mean value of $\bar{n}_2w_{\mathrm{p}}(r_{\mathrm{p}})$ can be obtained by:
\begin{equation}
\bar{n}_{2} w_{\mathrm{p}}\left(r_{\mathrm{p}}\right) = \frac{1}{N_{\mathrm{sub}}} \sum_{k=1}^{N_{\mathrm{sub}}} \bar{n}_{2, k} w_{\mathrm{p}, k}\left(r_{\mathrm{p}}\right).
\end{equation}
The corresponding error is calculated as:
\begin{equation}
\sigma^{2} = \frac{N_{\text{sub}}-1}{N_{\text{sub}}} \sum_{k=1}^{N_{\text{sub}}} \left(\bar{n}_{2, k} w_{\mathrm{p}, k}\left(r_{\mathrm{p}}\right) - \bar{n}_{2} w_{\mathrm{p}}\left(r_{\mathrm{p}}\right)\right)^{2}.
\end{equation}
Here, $\bar{n}_{2, k} w_{\mathrm{p}, k}\left(r_{\mathrm{p}}\right)$ indicates the excess of the projected density for the $k$th realization, and $N_{\mathrm{sub}}$ signifies the number of jackknife realizations. We use $N_{\mathrm{sub}}=50$ in this work. The measurements are displayed in Figure \ref{fit_nwp} as dots with error bars.

Although the SDSS Main sample includes lower mass galaxies, we only use centrals with \(M_* > 10^{11.1} M_{\odot}\) due to challenges in selecting central galaxies. However, while we do not use lower mass spectroscopic galaxies to build our 3D quenched fraction model, they serve as a useful tool to verify the extrapolation of our model. Therefore, we also measure the \(\bar{n}_2 w_{\mathrm{p}}(r_{\mathrm{p}})\) of companion galaxies with different colors and stellar masses around the entire lower mass spectroscopic sample, including both central and satellite galaxies. These measurements are then compared to the extrapolated predictions from our 3D quenched fraction model. We believe this approach provides strong validation for the extrapolability of our model.

\begin{figure*}[htb!]
\begin{center}
\includegraphics[width=0.8\textwidth]{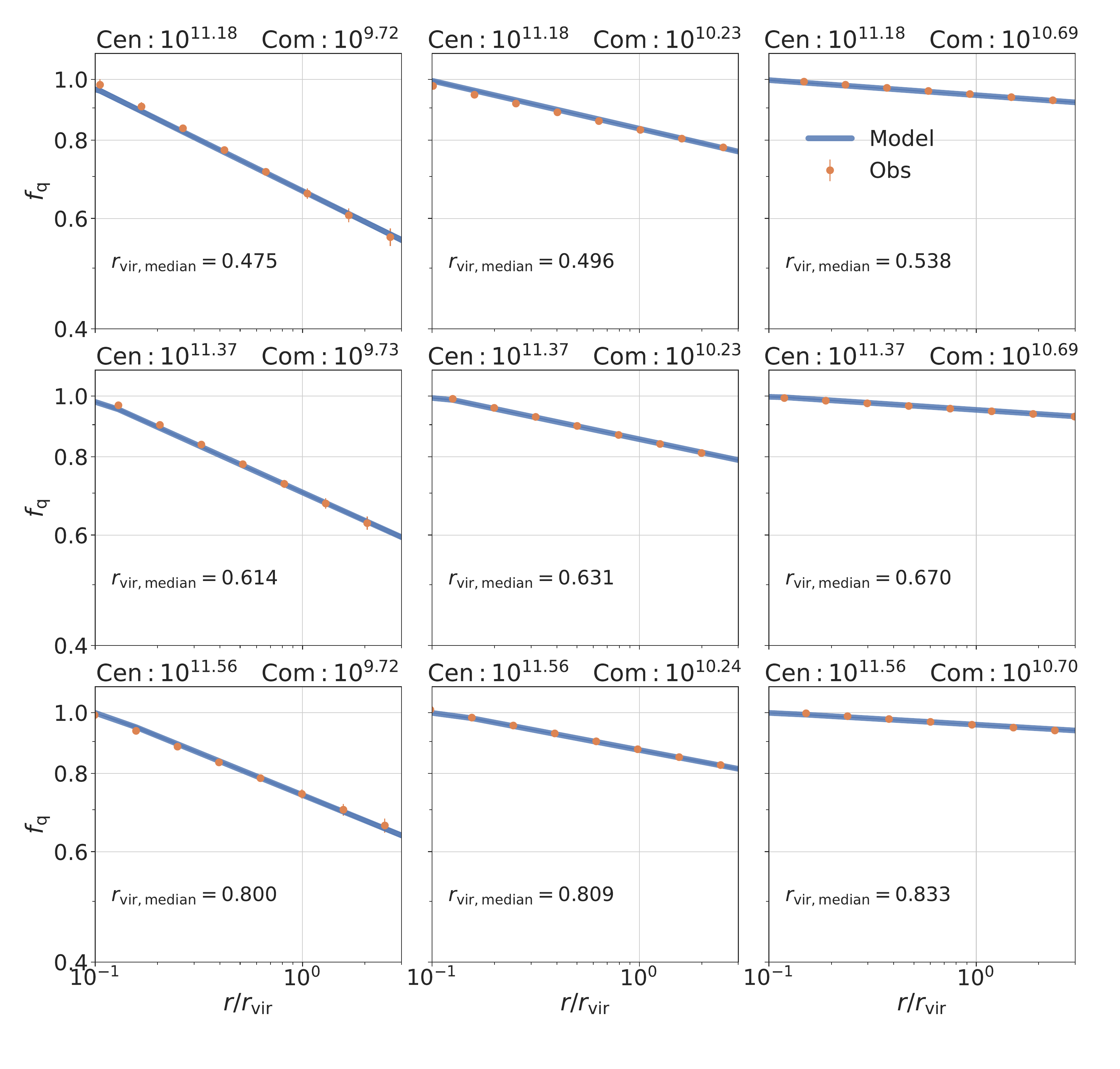}
\end{center}
\caption{The results of \(f^{\rm{sat}}_{\mathrm{q}}(r; M_{*,\rm{cen}}, M_{*,\rm{sat}})\) and \(f^{\rm{env}}_{\mathrm{q}}(r; M_{*,\rm{cen}}, M_{*,\rm{sat}})\) as functions of \(r/r_{\mathrm{vir}}\). The arrangement of subplots by mass bins follows the same format as in Figure \ref{fit_nwp}. Orange dots with error bars represent the observational results of \(f^{\rm{sat}}_{\mathrm{q}}\), calculated using Equation \ref{fq_def}. Blue lines correspond to \(f^{\rm{com}}_{\mathrm{q}}\) derived from the {\tt{Jiutian}} simulation after applying the 3D quenched fraction model (Equation \ref{fq_fit}). Green lines indicate \(f^{\rm{env}}_{\mathrm{q}}(r; M_{*,\rm{cen}}, M_{*,\rm{sat}})\) from {\tt{Jiutian}} simulation, calculated as \(f^{\rm{sat}}_{\mathrm{q}}(r; M_{*,\rm{cen}}, M_{*,\rm{sat}}) - \bar{f}^{\rm{cen}}(M_{*,\rm{sat}})\).}
\label{3d_fq}
\end{figure*}

\begin{figure*}[htbp]
\begin{center}
\includegraphics[width=0.9\textwidth]{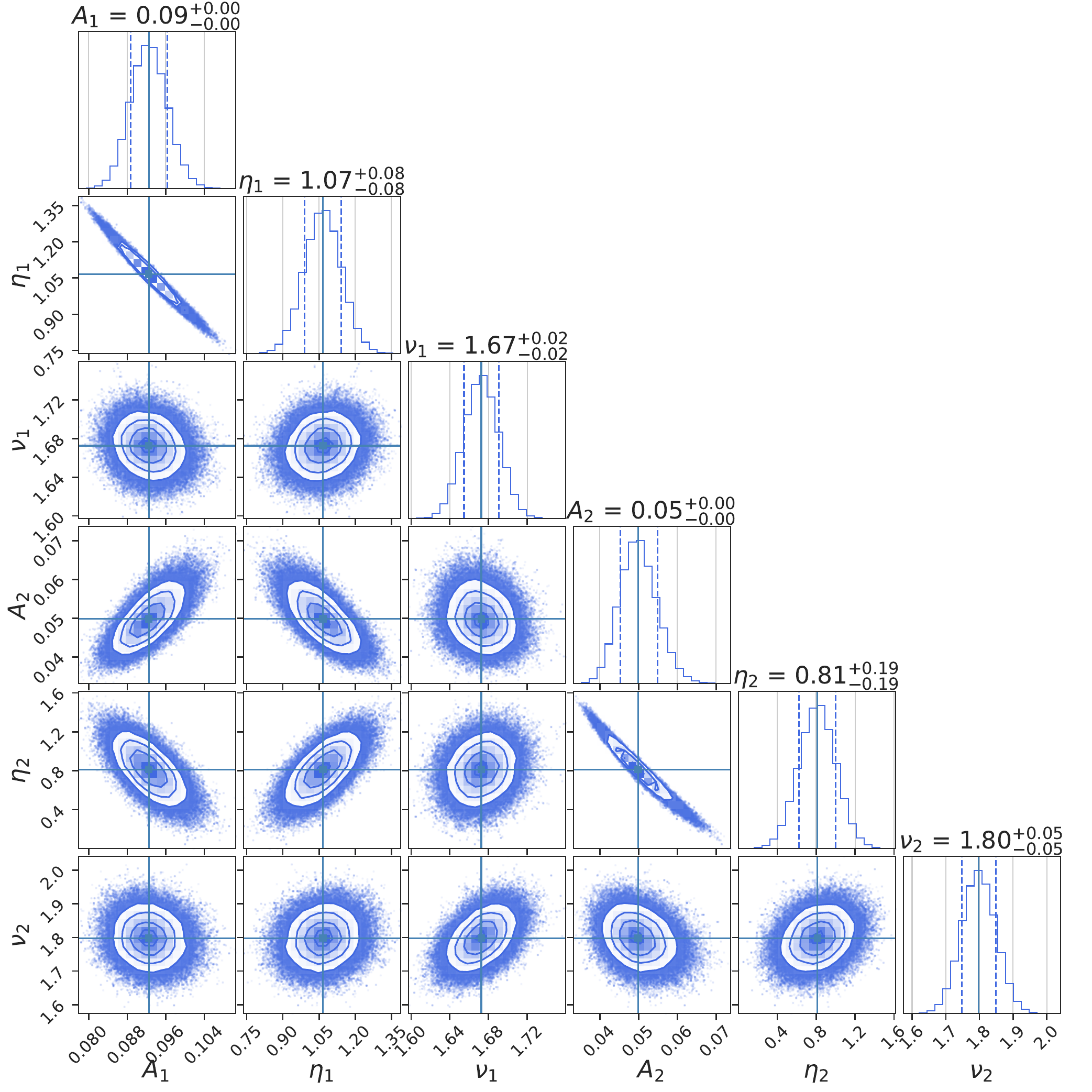}
\end{center}
\caption{The posterior distributions of the parameters in the fitting model for \(f_{\mathrm{q}}(r; M_{*,\rm{cen}}, M_{*,\rm{com}})\). The 1D PDF of each parameter is plotted as a histogram at the top panel of each column, where the median value and $1\sigma$ uncertainty is also labeled. The 2D joint PDF of each parameter pair is shown as a contour with three confidence levels $68\%$, $95\%$ and $99\%$.}
\label{MCMC_plot}
\end{figure*}

\section{Building the 3D quenched fraction Model} \label{sec:model}

In this section, we develop the 3D quenched fraction model by combining the measured \(\bar{n}_2 w_{\mathrm{p}}(r_{\mathrm{p}})\), the mean quenched fraction \(\bar{f}_{\rm{q}}^{\rm{all}}\), and the SHMR from the N-body simulation.

\subsection{3D quenched fraction profiles of satellite galaxies}
To derive the 3D quenched fraction profile, we first recover the 3D distributions of the red and the entire companion galaxy sample around massive central galaxies from the projected distributions \(\bar{n}_2 w_{\mathrm{p}}(r_{\mathrm{p}})\). Assuming that the real space correlation function $\xi(r)$ follows the power-law form $\xi(r) = (r/r_0)^{-\gamma}$, we have:
\begin{equation}
    w_{\mathrm{p}}\left(r_{\mathrm{p}}\right)=r_{\mathrm{p}}\left(\frac{r_{\mathrm{p}}}{r_{0}}\right)^{-\gamma} \Gamma\left(\frac{1}{2}\right) \Gamma\left(\frac{\gamma-1}{2}\right) / \Gamma\left(\frac{\gamma}{2}\right),
    \label{wp}
\end{equation}
where $\Gamma$ is the Gamma function. The power-law assumption is a good approximation for the scales of $< 3r_{\mathrm{vir}}$ that are relevant to this paper below. Then, \(\bar{n}_2 w_{\mathrm{p}}(r_{\mathrm{p}})\) can be written as:
\begin{equation}
\begin{aligned}
        \bar n_2w_{\mathrm{p}}\left(r_{\mathrm{p}}\right)&= \bar n_2 \left(r_0\right)^{\gamma}\left({r_{\mathrm{p}}}\right)^{1-\gamma} \Gamma\left(\frac{1}{2}\right) \Gamma\left(\frac{\gamma-1}{2}\right) / \Gamma\left(\frac{\gamma}{2}\right)\ \\
    &=\left(r_1\right)^{\gamma}\left({r_{\mathrm{p}}}\right)^{1-\gamma} \Gamma\left(\frac{1}{2}\right) \Gamma\left(\frac{\gamma-1}{2}\right) / \Gamma\left(\frac{\gamma}{2}\right),
\end{aligned}
\end{equation}
where we absorb $n_2$ and $r_0$ into $r_1$. By fitting $\bar{n}_{2} w_{\mathrm{p}}\left(r_{\mathrm{p}}\right)$ of companion galaxies in the range $0.1<r_{\mathrm{p}}<3 h^{-1}\mathrm{Mpc}$, we can constrain $r_1$ and $\gamma$. We show the fitting results in Figure \ref{fit_nwp} with lines. It is evident that the data and fit exhibit a relatively good agreement across the whole mass range, both for the entire galaxy samples and for the red ones.

\begin{figure}[htb!]
\includegraphics[width=0.45\textwidth]{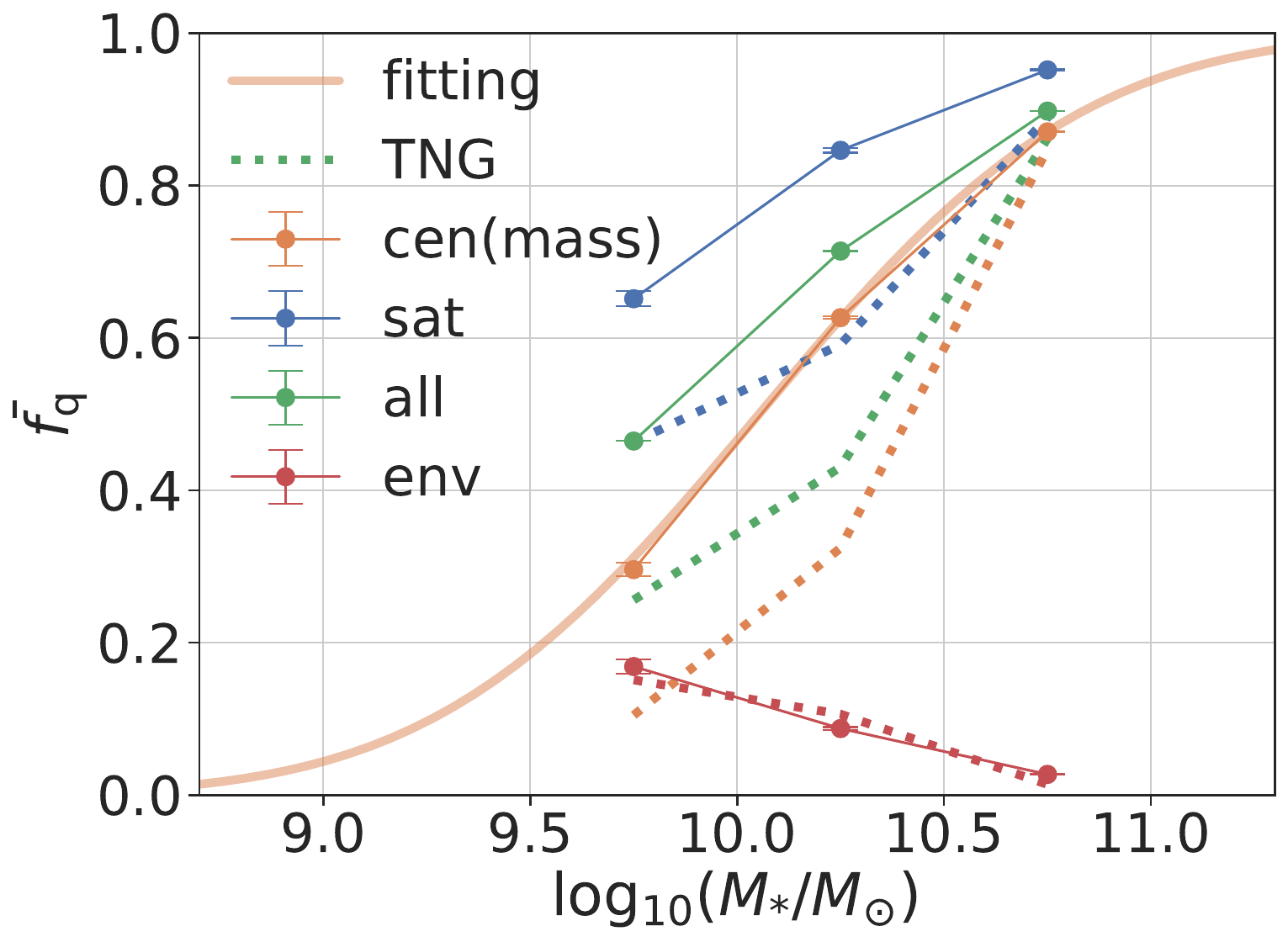}
\caption{The average quenched fraction of satellite (blue), central (orange), and all galaxies (green) as a function of stellar mass. The environmental quenched fraction (red) is also shown, defined as \(f^{\rm{env}}(M_{*}) = f^{\rm{all}}(M_{*}) - f^{\rm{cen}}(M_{*})\). The orange line represents the fitted trend for \(f^{\rm{cen}}(M_{*})\).  For comparison, the corresponding quenched fractions from {\texttt{TNG300-1}} are also included.}
\label{central_fq}
\end{figure}

\begin{figure*}[!htp]
\begin{center}
\includegraphics[width=0.8\textwidth]{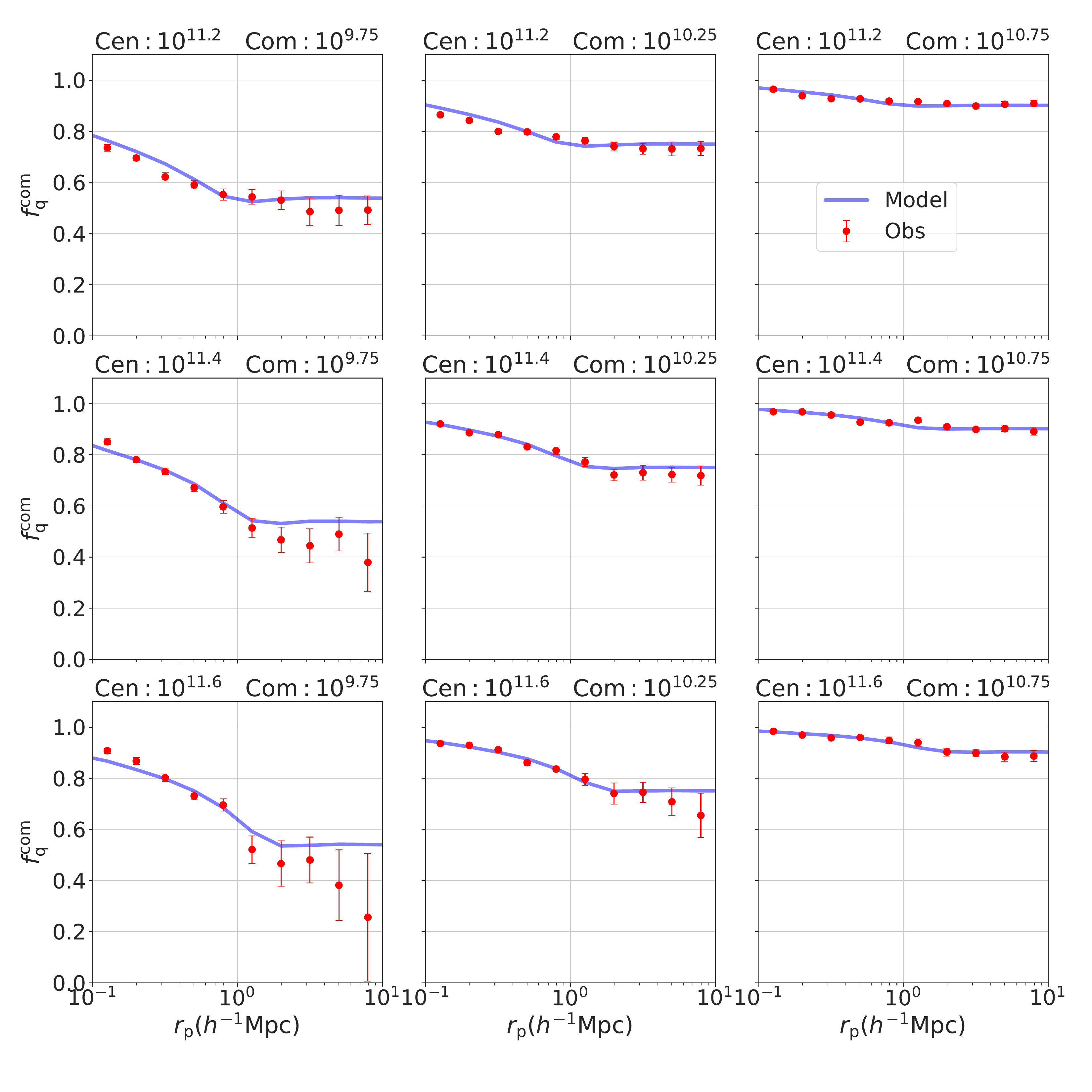}
\end{center}
\caption{The projected quenched fraction  \(f^{\rm{com}}_{\rm{q}}(r_{\rm{p}};M_{*,{\rm{cen}}},M_{*,\rm{com}})\) of companion galaxies around central galaxies is represented by red dots with error bars for the observational data and blue lines for the simulation. The arrangement of subplots by mass bins follows the same format as in Figure \ref{fit_nwp}.}
\label{cen_frac}
\end{figure*}

\begin{figure*}
\begin{center}
\includegraphics[width=0.65\textwidth]{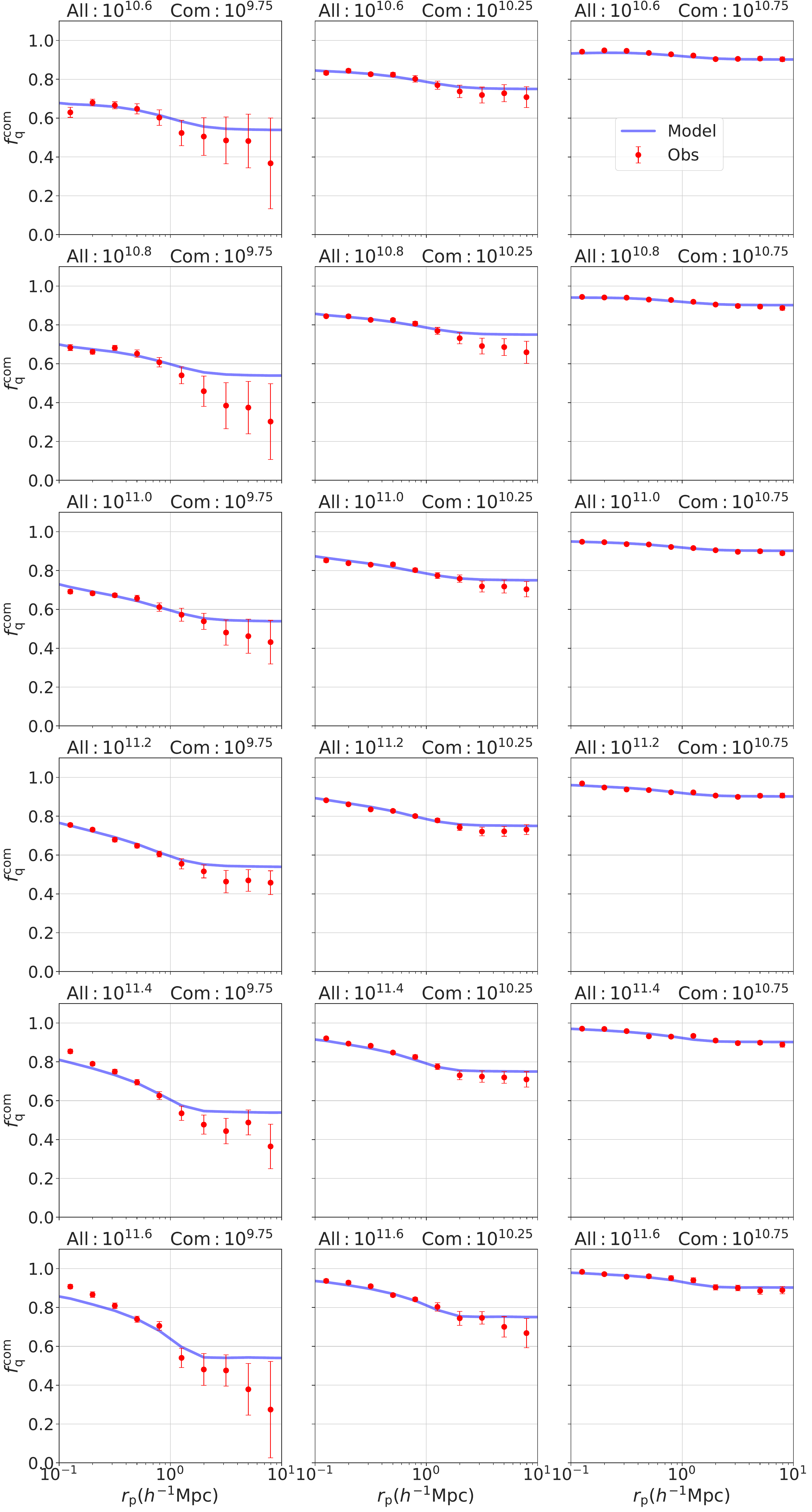}
\end{center}
\caption{The projected quenched fraction \(f^{\rm{com}}_{\rm{q}}(r_{\rm{p}};M_{*,{\rm{all}}},M_{*,\rm{com}})\) of companion galaxies around all spectroscopic galaxies for observational data and simulation, shown as red dots with error bars and blue lines respectively. From left to right, mass bins for the companion samples vary from \(10^{9.2}\) to \(10^{11.0} M_{\odot}\). From top to bottom, the spectroscopic stellar mass bins change from \(10^{10.5}\) to \(10^{11.7} M_{\odot}\). The subplot titles indicate the corresponding bin centers.}
\label{check_frac}
\end{figure*}

Consequently, the 3D number density distribution can be writen as:
\begin{equation}
    \bar n_2\xi(r)=\bar n_2\left(r / r_0\right)^{-\gamma}  = (r_1)^{\gamma}(r)^{-\gamma}.
\end{equation}

Similarly, for red sub-samples, we can obtain $\bar{n}_{2}^{red} \xi^{red}\left(r\right)$. Finally, we can compute the 3D quenched fraction distributions for companion galaxies:
\begin{equation}
    f^{\rm{com}}_{\mathrm{q}}(r) = \frac{\bar{n}_{2}^{red} \xi^{red}\left(r\right)}{\bar{n}_{2} \xi\left(r\right)}
    = \frac{(r_1^{red})^{{\gamma}^{red}}r^{{-\gamma}^{red}}}{(r_1)^{\gamma}r^{-\gamma}},
    \label{fq_def}
\end{equation}

In Figure \ref{3d_fq}, we present the 3D quenched fraction distributions. The mass bins for central and companion galaxies are consistent with those shown in Figure \ref{fit_nwp}. For each bin, we list the median values of central stellar mass, companion stellar mass and $r_{\mathrm{vir}}$ derived from the {\tt{Jiutian}} catalog, which are subsequently used to build the model. It is clear that the fraction $f^{\rm{com}}_{\mathrm{q}}$ declines as the distance $r/r_{\mathrm{vir}}$ increases, following a power-law relation. This indicates that galaxies located closer to central galaxies are more likely to be quenched, thereby confirming the environmental effect. Additionally, we observe a mass dependence of \(f^{\rm{com}}_{\mathrm{q}}\): Higher masses correspond to higher \(f^{\rm{com}}_{\mathrm{q}}\), a trend evident for both central and companion stellar masses. The monotonous increase with companion mass highlights the impact of mass quenching, while the gradual increase with central mass suggests a stronger environmental effect in more massive halos, even when scaled distances \(r/r_{\text{vir}}\) are taken into account.

Based on these observational results, we construct a fitting formula of $f^{\rm{com}}_{\mathrm{q}}$ as follows:
\begin{equation}
f^{\rm{com}}_{\text{q}} (r;M_{*,\rm{cen}},M_{*,\rm{com}}) = A \left(\frac{r}{r_{\text{vir}}}\right)^{\kappa}, 
\label{fq_fit}
\end{equation}
where
\begin{equation}
\begin{aligned}
&A =  \\
& 1 - A_1 * (13-log_{10}(M_{*,{\mathrm{cen}}}))^{\eta_1} * (11.2-log_{10}(M_{*,{\mathrm{com}}}))^{\nu_1},
\nonumber
\end{aligned}
\end{equation}

\begin{equation}
\begin{aligned}
&{\kappa} =  \\
& - A_2 * (13-log_{10}(M_{*,{\mathrm{cen}}}))^{\eta_2} *(11.2-log_{10}(M_{*,{\mathrm{com}}}))^{\nu_2},
\nonumber
\end{aligned}
\end{equation}
where $A_1, \eta_1, \nu_1$ and $A_2, \eta_2, \nu_2$ are free parameters of the model. In order to fit the parameters, we employ the Markov Chain Monte Carlo (MCMC) sampler \texttt{emcee} \citep{2013PASP..125..306F}. We use the median values of central stellar mass, companion stellar mass, and $r_{\rm{vir}}$ for each mass bin, as previously mentioned, in the fitting process. The posterior probability density functions (PDFs) of the parameters are shown in Figure \ref{MCMC_plot}, from which we can conclude that all parameters are well constrained. To validate the fitting, we compare it to the observed \(f^{\rm{com}}_{\mathrm{q}}(r;M_{*,\rm{cen}},M_{*,\rm{com}})\) in Figure \ref{3d_fq}. Overall, the fits of \(f^{\rm{com}}_{\mathrm{q}}\) are consistent with the observational data within the error margins. 

Although Equation \ref{fq_fit} provides \(f^{\rm{com}}_{\mathrm{q}}(r;M_{*,\rm{cen}},M_{*,\rm{com}})\) to large scales, in our model, we only apply it to satellite galaxies within the radius \(r_{\rm{FOF}}\) of each FOF halo. We define the 3D satellite quenched fraction profile as 

\begin{align}
    &f^{\rm{sat}}_{\mathrm{q}}(r;M_{*,\rm{cen}},M_{*,\rm{sat}})\notag\\
    &=f^{\rm{com}}_{\mathrm{q}}(r;M_{*,\rm{cen}},M_{*,\rm{sat}}), \ 0<r<r_{\rm{FOF}}.
\end{align}

We use \(f_q^{\rm{sat}}\) instead of \(f_q^{\rm{com}}\) to ensure the uniqueness of the model prediction, as each satellite galaxy is associated with only one central galaxy, whereas companion galaxies outside \(r_{\rm{FOF}}\) can be linked to multiple central galaxies. Moreover, this approach is more physically motivated, as \citetalias{2024ApJ...969..129Z} demonstrated that environmental quenching predominantly occurs within halos, while companion galaxies beyond \(r_{\rm{FOF}}\) are primarily influenced by their own host halos.

\subsection{Quenched fraction of central galaxies}
With the 3D quenched fraction model for satellite galaxies developed above, we can determine the quenched fraction of each satellite galaxy. However, to complete the picture, we still need the quenched fraction of central galaxies, \(\bar{f}^{\rm{cen}}_{\rm{q}}(M_*)\). Additionally, since the quenching of satellite galaxies results from both mass and environmental effects, \(\bar{f}^{\rm{cen}}_{\rm{q}}(M_{*})\) is crucial for disentangling these two factors, as it is believed to represent the mass-driven effect.

While we could derive  \(\bar{f}^{\rm{cen}}_{\rm{q}}(M_{*})\) using the existing \(f^{\rm{com}}_{\mathrm{q}}(r; M_{*,\rm{cen}}, M_{*,\rm{com}})\) results beyond \(r_{\rm{vir}}\) with iterative methods, it would be much simpler to include a new observational quantity, \(\bar{f}^{\rm{all}}_{\rm{q}}(M_{*})\), the average quenched fraction for all galaxies in each stellar mass bin. \(\bar{f}^{\rm{all}}_{\rm{q}}(M_{*})\) can be easily obtained by counting blue and red galaxies in each stellar mass bin. To account for completeness, during the counting, we employ a weight of $1/V_{\mathrm{max}}$ for each galaxy, where $V_{\mathrm{max}}$ represents the volume corresponding to $z_{\mathrm{max}}$, the maximum redshift at which the galaxy satisfies our selection criteria. The corresponding errors are estimated from the bootstrap resamplings. With \(\bar{f}^{\rm{all}}_{\rm{q}}(M_{*})\) and the average quenched fraction of satellite galaxies, \(\bar{f}^{\rm{sat}}_{\rm{q}}(M_{*})\), calculated from our model, we can solve for \(\bar{f}^{\rm{cen}}_{\rm{q}}(M_{*})\) :
\begin{equation}
    {\bar{f}_{\mathrm{q}}}^{\mathrm{cen}} = \frac{\bar{f}_{\mathrm{q}}^{\rm{all}} * (N_{\mathrm{sat}}+N_{\mathrm{cen}}) - N_{\mathrm{sat}}*\bar{f}_{\mathrm{q}}^{\rm{sat}}}{N_{\mathrm{cen}}}.
    \label{eq:fq_cen}
\end{equation}
Here, \(N_{\mathrm{sat}}\) denotes the number of satellite galaxies in each stellar mass bin, while \(N_{\mathrm{cen}}\) represents the number of central galaxies. These values are calculated from our models using the {\tt{Jiutian}} simulation. We would like to note that in the calculation of \(\bar{f}^{\rm{sat}}(M_{*})\), we have extrapolated our \(f^{\rm{sat}}_{\mathrm{q}}(r; M_{*,\rm{cen}}, M_{*,\rm{sat}})\) model. This is because the calculation requires the \(f^{\rm{sat}}\) values for all satellite galaxies around central galaxies with any stellar mass, while the model is built based on measurements for central galaxies only with \(M_{*,{\rm{cen}}}>10^{11.1}M_{\odot}\). This extrapolation will be validated in Section \ref{sec:validation} by incorporating additional measurements. For now, we assume the extrapolation is reasonable.

\begin{figure*}
\centering
\includegraphics[width=0.7\textwidth]{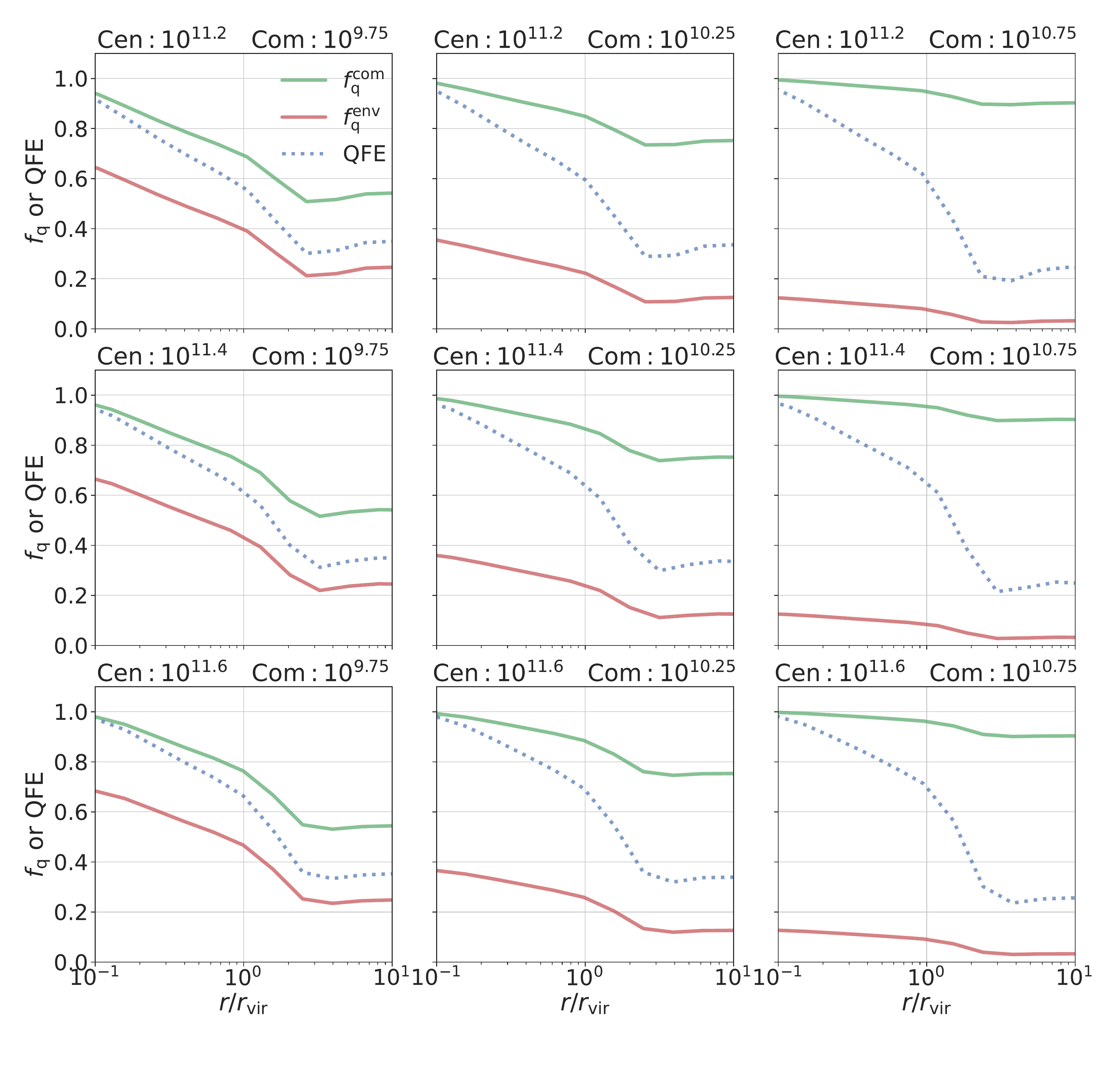}
\caption{The total \(f^{\rm{com}}_{\rm{q}}(r;M_{*,{\rm{cen}}},M_{*,\rm{com}})\) and environmental \(f^{\rm{env}}_{\rm{q}}(r;M_{*,{\rm{cen}}},M_{*,\rm{com}})\) quenched fractions of companion galaxies around central galaxies derived from our model. The arrangement of subplots by mass bins follows the same format as in Figure \ref{3d_fq}.}
\label{fig:radius_frac}
\end{figure*}

In Figure \ref{central_fq}, we show the derived \(\bar{f}^{\rm{sat}}(M_{*})\), \(\bar{f}^{\rm{cen}}(M_{*})\), \(\bar{f}^{\rm{all}}(M_{*})\), and the environmental quenched fraction \(\bar{f}^{\rm{env}}(M_{*}) = \bar{f}^{\rm{all}}(M_{*}) - \bar{f}^{\rm{cen}}(M_{*})\), where \(\bar{f}^{\rm{cen}}(M_{*})\) also represents the mass quenched fraction, \(\bar{f}^{\rm{mass}}(M_{*})\), in our definition. We find that \(\bar{f}^{\rm{cen}}\) increases monotonically with stellar mass, suggesting that mass quenching effects become more pronounced in more massive galaxies.To describe \(\bar{f}^{\rm{cen}}(M_{*})\), we construct a model as follows:

\begin{equation}
    \bar{f}^{\rm{cen}}_{\mathrm{q}}(M_*)=\frac{1}{2}\left(1+\operatorname{erf}\left(\frac{\log _{10}\left(M_{\mathrm{*}}\right)-\mu}{\sqrt{2} \sigma }\right)\right).\label{eq:fq_cen_fit}
\end{equation}

Here, \(\operatorname{erf}\) is the error function, \(\mu\) is the central location, and \(\sigma\) is the width of the distribution. The fitted values for \(\mu\) and \(\sigma\) are 10.02 and 0.64, respectively. The fitting model is displayed in Figure \ref{central_fq} with orange line. Using this model, we can assign the value of \(\bar{f}_{\mathrm{q}}^{\mathrm{cen}}\) to each central galaxy in the {\tt{Jiutian}} simulation. Combined with the \(f^{\rm{sat}}_{\rm{q}}\) model, we can now determine the quenched probability of each galaxy in the simulation. We present the model prediction of \(f^{\rm{com}}_{\mathrm{q}}(r;M_{*,\rm{cen}},M_{*,\rm{com}})\) from the {\tt{Jiutian}} simulation in Figure \ref{3d_fq} and find that our model accurately reproduces the measurements.

\subsection{Validating the Model Extrapolation}\label{sec:validation}
The 3D quenched fraction model we developed in Equation \ref{fq_fit} and Equation \ref{eq:fq_cen_fit} can predict the quenched probability for any galaxy with \(M_* > 10^{9.5}M_{\odot}\). However, as mentioned above, this is achieved by extrapolating the \(f^{\rm{sat}}_{\mathrm{q}}(r; M_{*,\rm{cen}}, M_{*,\rm{sat}})\) model to \(M_{*,{\rm{cen}}} < 10^{11.1}M_{\odot}\), and this extrapolation still needs to be validated.

Before validate the extrapolation, we first check if we can reproduce the measurements with \(M_{*,{\rm{cen}}} > 10^{11.1}M_{\odot}\) used to construct our model. The measurements we choose to compare is the projected quenched fraction of companion galaxies $f^{\rm{com}}_{\rm{q}}(r_{\rm{p}};M_{*,{\rm{cen}}},M_{*,\rm{com}})$ as used in \citetalias{2024ApJ...969..129Z}. The $f^{\rm{com}}_{\rm{q}}(r_{\rm{p}};M_{*,{\rm{cen}}},M_{*,\rm{com}})$ of the \(k\)th jackknifed realization is 
\begin{equation}
    f^{\rm{com}}_{\mathrm{q}, k}(r_{\mathrm{p}}) = \frac{\bar{n}_{2, k}^{red} w_{\mathrm{p}, k}^{red}\left(r_{\mathrm{p}}\right)}{\bar{n}_{2, k} w_{\mathrm{p}, k}\left(r_{\mathrm{p}}\right)}.
\end{equation}
Then, the mean value of the $f^{\rm{com}}_{\rm{q}}(r_{\rm{p}};M_{*,{\rm{cen}}},M_{*,\rm{com}})$ and the corresponding error can be expressed as:
\begin{equation}
    f^{\rm{com}}_{\mathrm{q}}(r_{\mathrm{p}}) = \frac{1}{N_{\mathrm{sub}}} \sum_{k=1}^{N_{\mathrm{sub}}} f^{\rm{com}}_{\mathrm{q}, k}(r_{\mathrm{p}})\,\,,
    \label{eq:jack_fq}
\end{equation}
\begin{equation}
    \sigma^{2}_{f^{\rm{com}}_{\mathrm{q}}}(r_{\mathrm{p}}) = \frac{N_{\text {sub }}-1}{N_{\text {sub }}} \sum_{k=1}^{N_{\text {sub }}}\left(f^{\rm{com}}_{\mathrm{q}, k}(r_{\mathrm{p}})- f^{\rm{com}}_{\mathrm{q}}(r_{\mathrm{p}}) \right)^{2} \,\,.
    \label{eq:jack_error}
\end{equation}
We note that the observed \(f^{\rm{com}}_{\rm{q}}(r_{\rm{p}};M_{*,{\rm{cen}}},M_{*,\rm{com}})\) is a direct measurement from observations, independent of any model assumptions.

In Figure \ref{cen_frac}, we compare the observed \(f^{\rm{com}}_{\rm{q}}(r_{\rm{p}};M_{*,{\rm{cen}}},M_{*,\rm{com}})\) with predictions from our model across all stellar mass bins used to construct the model, spanning scales of \(0.1 < r_{\rm{p}} < 10 \, h^{-1}\,\mathrm{Mpc}\). We find that our model accurately reproduces the measurements across all stellar mass bins, confirming its self-consistency. This test also provides a weak validation of the extrapolation, as \(f^{\rm{com}}_{\rm{q}}(r_{\rm{p}})\), particularly at \(r_{\rm{p}} > r_{\rm{vir}}\), contains some information about the quenched fraction of satellite galaxies with \(M_{*,\rm{cen}} < 10^{11.0}M_{\odot}\). This is because companion galaxies at \(r > r_{\rm{vir}}\) have the potential to be satellite galaxies of these less massive centrals, contributing to \(f^{\rm{com}}_{\rm{q}}(r_{\rm{p}})\). However, since the measurements at \(r_{\rm{p}} > r_{\rm{vir}}\) have larger uncertainties, this serves only as a relatively weak validation.

To further validate the extrapolation, we incorporate additional measurements as described below. The reason we initially used only measurements with \(M_{*,{\rm{cen}}} > 10^{11.1}M_{\odot}\) to construct the model is that identifying lower-mass central galaxies in observations is both challenging and prone to significant uncertainties. However, in the model, we can easily predict \(f^{\rm{com}}_{\rm{q}}(r_{\rm{p}};M_{*,{\rm{all}}},M_{*,\rm{com}})\), which represents the quenched fraction of companion galaxies around all galaxies with \(M_{*,{\rm{all}}}\), including both central and satellite galaxies. This eliminates the need to isolate central galaxies in observations for comparison. This approach allows us to extend the measurements of \(f^{\rm{com}}_{\rm{q}}(r_{\rm{p}};M_{*,{\rm{all}}},M_{*,\rm{com}})\) to \(M_{*,{\rm{all}}} < 10^{11.1}M_{\odot}\) using the SDSS spectroscopic sample, providing a stronger test for our model.

In Figure \ref{check_frac}, we compare the measured and predicted \(f^{\rm{com}}_{\rm{q}}(r_{\rm{p}};M_{*,{\rm{all}}},M_{*,\rm{com}})\) for \(M_{*,{\rm{all}}} > 10^{10.5}M_{\odot}\) and within the same \(M_{*,\rm{com}}\) ranges. We refrain from extending to lower \(M_{*,{\rm{all}}}\) because the SDSS is only complete at very low redshifts, and the PAC method used in this study has not been fully validated at such low redshifts. This extension can be achieved with next-generation spectroscopic surveys and an improved PAC method (K. Xu et al., in prep.). We find that our model accurately reproduces all the \(f^{\rm{com}}_{\rm{q}}(r_{\rm{p}};M_{*,{\rm{all}}},M_{*,\rm{com}})\) measurements down to \(M_{*,{\rm{all}}} > 10^{10.5}M_{\odot}\), providing a strong test of our model in the environments of $M_{\rm{vir}}>10^{12.0}h^{-1}M_{\odot}$.

 \begin{figure*}
\begin{center}
\includegraphics[width=1\textwidth]{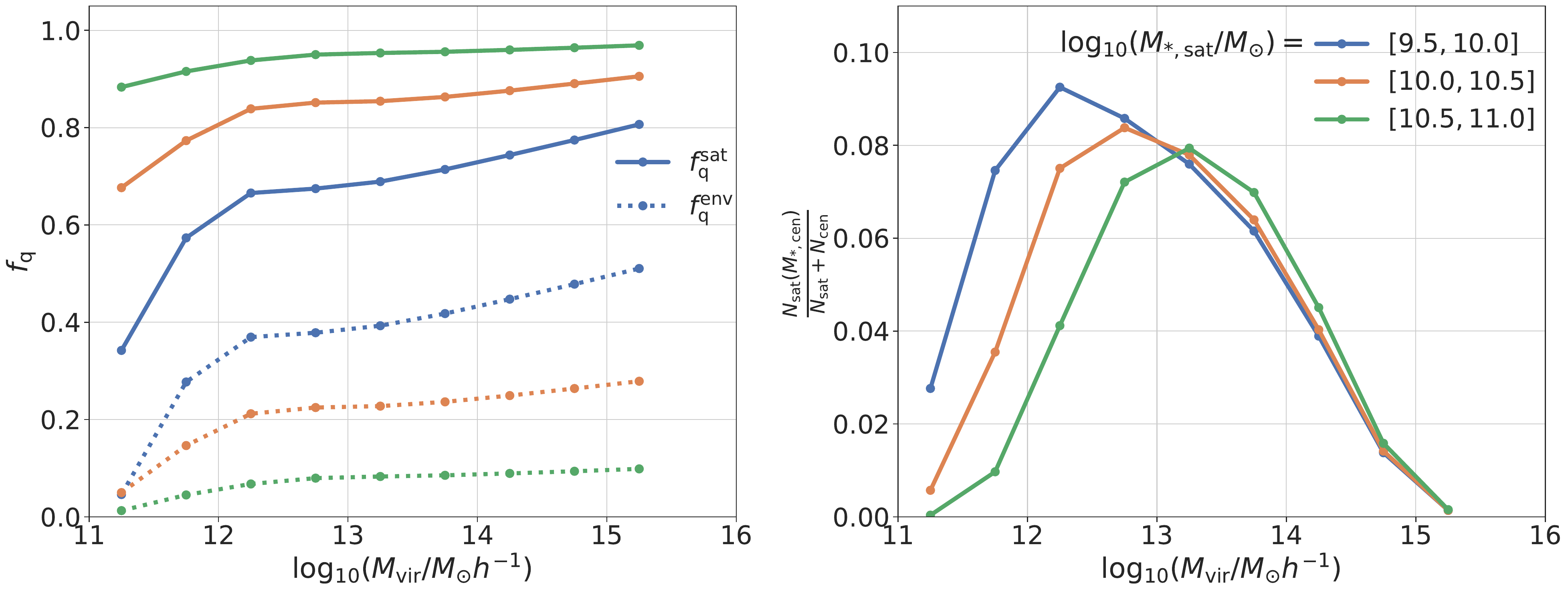}
\end{center}
\caption{Left: The total quenched fraction (mass+environment), \(f^{\rm{sat}}_{\mathrm{q}}(M_{*,\rm{cen}},M_{*,\rm{sat}})\), and the environment quenched fraction, \(f^{\rm{env}}_{\mathrm{q}}(M_{*,\rm{cen}},M_{*,\rm{sat}})\), of satellite galaxies as functions of stellar mass. Right: The fraction of galaxies, in each stellar mass bin, that are satellite galaxies associated with halos of specific halo masses.}
\label{fig:total_fq}
\end{figure*}

\begin{figure*}
\begin{center}
\includegraphics[width=1\textwidth]{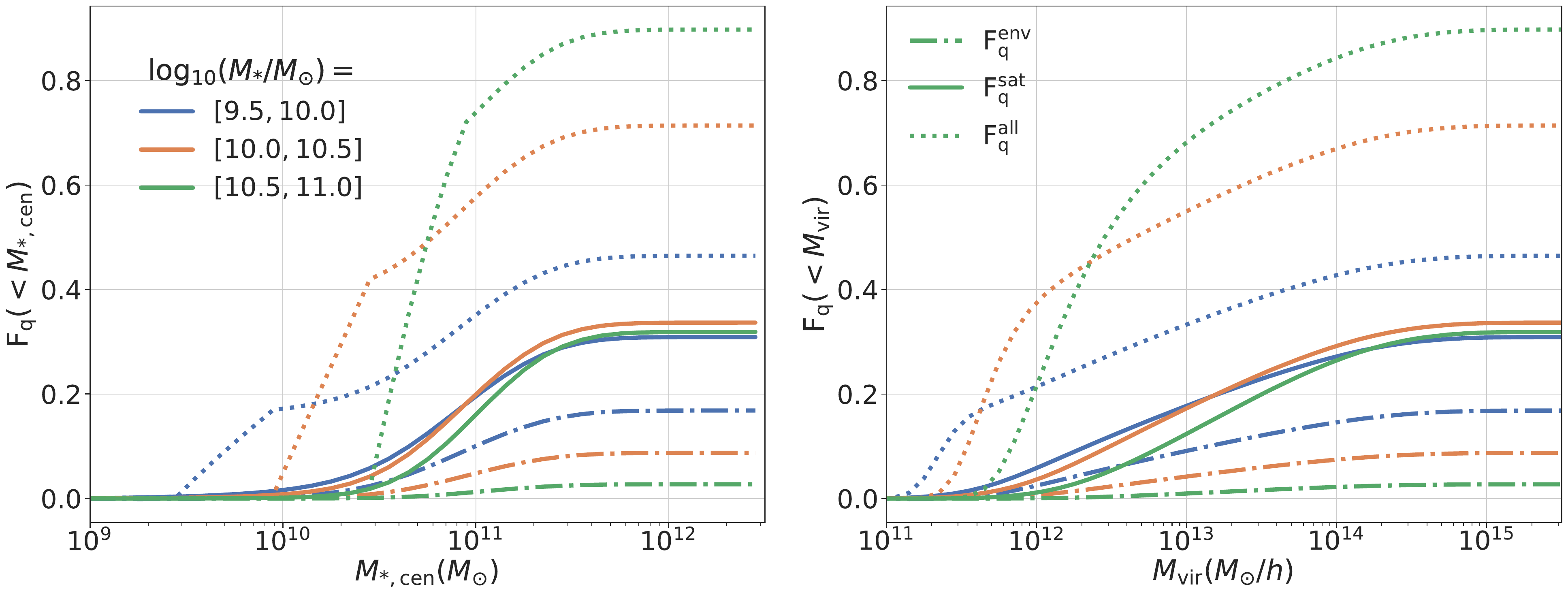}
\end{center}
\caption{The cumulative quenched fraction from environment (dashed lines), for satellite galaxies (solid lines), and for all the galaxies (dotted lines) as a function of central galaxy mass (left) and host halo mass (right).}
\label{all_fq}
\end{figure*}

\section{Quenching effects of mass and environment}\label{sec:quenching}
With our 3D quenched fraction model, \(f^{\rm{sat}}_{\mathrm{q}}(r; M_{*,\rm{cen}}, M_{*,\rm{sat}})\), \(\bar{f}^{\rm{cen}}_{\rm{q}}(M_{*})\), and the galaxy-halo connection in N-body simulations, we can now disentangle the effects of mass and environment on galaxy quenching.

\subsection{Observational results}
In Figure \ref{fig:radius_frac}, we present the 3D quenched fraction of companion galaxies, \(f^{\rm{com}}_{\rm{q}}(r;M_{*,{\rm{cen}}},M_{*,\rm{com}})\), and the environmental contribution, \(f^{\rm{env}}_{\rm{q}}(r;M_{*,{\rm{cen}}},M_{*,\rm{com}}) = f^{\rm{com}}_{\rm{q}}(r;M_{*,{\rm{cen}}},M_{*,\rm{com}}) - \bar{f}^{\rm{cen}}_{\rm{q}}(M_{*,\rm{com}})\). This figure is similar to Figure \ref{3d_fq} but extends the analysis to \(10r/r_{\rm{vir}}\). We find that both \(f^{\rm{com}}_{\rm{q}}\) and \(f^{\rm{env}}_{\rm{q}}\) decrease with \(r\) within \(2-3r_{\rm{vir}}\), with more rapid changes occurring at \(r > r_{\rm{vir}}\) compared to \(r < r_{\rm{vir}}\). Beyond \(3r_{\rm{vir}}\), \(f^{\rm{com}}_{\rm{q}}\) and \(f^{\rm{env}}_{\rm{q}}\) remain nearly constant, reflecting the average total and environmental quenched fractions. We suggest that the constant quenched fraction observed at \(r > 3r_{\rm{vir}}\), rather than \(r > r_{\rm{vir}}\), is due to the scatter in \(r_{\rm{vir}}\) at fixed central stellar mass and the triaxial shapes of halos \citep{2002ApJ...574..538J}, which allow the host halo environment to still influence galaxies beyond \(r_{\rm{vir}}\). The more rapid decrease in \(f^{\rm{com}}_{\rm{q}}\) and \(f^{\rm{env}}_{\rm{q}}\) beyond \(r_{\rm{vir}}\) can also be attributed to this, as only a subset of host halos exert influence at these distances.

From Figure \ref{fig:radius_frac}, it is evident that \(f^{\rm{env}}_{\rm{q}}\) increases with \(M_{*,{\rm{cen}}}\) and decreases with \(M_{*,{\rm{com}}}\). The dependence of \(f^{\rm{env}}_{\rm{q}}\) on \(M_{*,{\rm{com}}}\) may be attributed to the rapid decrease in the number of isolated blue galaxies that can be quenched by the environment, as \(\bar{f}^{\rm{cen}}_{\rm{q}}(M_{*,\rm{com}})\) increases sharply with \(M_{*,{\rm{com}}}\). To better describe the effects of environment, we define the 3D QFE:
\begin{equation}
    {\rm{QFE}}(r;M_{*,{\rm{cen}}},M_{*,\rm{com}}) =\frac{f^{\rm{env}}_{\rm{q}}(r;M_{*,{\rm{cen}}},M_{*,\rm{com}})}{1-\bar{f}^{\rm{cen}}_{\rm{q}}(M_{*,\rm{com}})} \,.\label{eq:QFE}
\end{equation}
The definition is similar to but different from that of \citetalias{2024ApJ...969..129Z}.
The QFE distributions for different central and companion stellar mass bins are shown as blue dotted lines in Figure \ref{fig:radius_frac}. We find that \(M_{*,{\rm{com}}}\) has only a very slight effect on QFE at \(r < r_{\rm{vir}}\), consistent with the conclusion of \citetalias{2024ApJ...969..129Z} that environmental quenching is largely independent of the companion mass. At \(r > r_{\rm{vir}}\), QFE decreases with \(M_{*,{\rm{com}}}\). We suspect this is because the average environment varies at \(r > r_{\rm{vir}}\) for different \(M_{*,{\rm{com}}}\), whereas at \(r < r_{\rm{vir}}\), the environment is determined by the fixed \(M_{*,{\rm{cen}}}\). Moreover, we find that QFE increases with \(M_{*,{\rm{cen}}}\), as expected, indicating that more massive host halos are more efficient at quenching their satellite galaxies.

To better quantify the fraction of satellite galaxies that are quenched in different environments, we plot the total $f_{\rm{q}}^{\rm{sat}}(M_{\rm{vir}},M_{*,{\rm{sat}}})$ and environmental $f_{\rm{q}}^{\rm{env}}(M_{\rm{vir}},M_{*,{\rm{sat}}})$ quenched fractions of satellite galaxies as a function of host halo mass in the left panel Figure \ref{fig:total_fq}. Similar to Figure \ref{fig:radius_frac}, we find that \(f_{\rm{q}}^{\rm{sat}}\) and \(f_{\rm{q}}^{\rm{env}}\) increase slightly with \(M_{\rm{vir}}\). \(f_{\rm{q}}^{\rm{sat}}\) increases rapidly with \(M_{*,{\rm{sat}}}\), while \(f_{\rm{q}}^{\rm{env}}\) decreases sharply with \(M_{*,{\rm{sat}}}\) due to mass quenching. Even in the most massive host halos, only 10\% of satellite galaxies in the \( [10^{10.5},10^{11.0}]M_{\odot} \) stellar mass bin are quenched due to environmental effects, while more than 50\% of those in the \( [10^{9.5},10^{10.0}]M_{\odot} \) stellar mass bin are. It is worth mentioning that \(f_{\rm{q}}^{\rm{sat}}\) and \(f_{\rm{q}}^{\rm{env}}\) increase rapidly for \(M_{\rm{vir}} < 10^{12.0} h^{-1} M_{\odot}\), which is primarily due to our parametrization of \(A\) and \(\kappa\) in Equation \ref{fq_fit}. These parameters depend on \(M_{*,{\rm{cen}}}\) rather than \(M_{\rm{vir}}\). As shown in Figure \ref{SHMR}, the SHMR exhibits a slope change around \(M_{\rm{vir}} = 10^{12.0} h^{-1} M_{\odot}\), which leads to a corresponding slope change in both \(f_{\rm{q}}^{\rm{sat}}(M_{\rm{vir}},M_{*,{\rm{sat}}})\) and \(f_{\rm{q}}^{\rm{env}}(M_{\rm{vir}},M_{*,{\rm{sat}}})\) at the same mass. Since our model has only been tested for \(M_{\rm{vir}} > 10^{12.0} h^{-1} M_{\odot}\) in Figure \ref{check_frac}, the behavior for \(M_{\rm{vir}} < 10^{12.0} h^{-1} M_{\odot}\) requires further investigation.

Finally, we examine total mass and environmental quenching within each stellar mass bin. These quantities depend on the host halo mass distribution of satellite galaxies, shown in the right panel of Figure \ref{fig:total_fq}. We then calculate the cumulative quenched fractions, \(F_{\rm{q}}\), due to the environment, for satellite galaxies, and for all galaxies, as a function of both central galaxy stellar mass and host halo mass, as presented in Figure \ref{all_fq}. The values at the highest mass end represent the total quenched fractions, which are also shown in Figure \ref{central_fq}. We find that the mass-quenched fraction increases from 0.3 to 0.87 across the stellar mass range \([10^{9.5}, 10^{10.0}]M_{\odot}\) to \([10^{10.5}, 10^{11.0}]M_{\odot}\), while the environmental quenched fraction decreases from 0.17 to 0.03. Mass effects dominate galaxy quenching across the entire stellar mass range of \([10^{9.5}, 10^{11.0}]M_{\odot}\), consistent with the findings of \cite{2019ApJ...882..167C}.

\subsection{Comparing to IllustrisTNG}
 \citetalias{2024ApJ...969..129Z} compared the quenched fraction derived from \(\bar{n}_2w_{\rm{p}}\) measurements to those from \texttt{TNG300-1} \citep{2018MNRAS.480.5113M, 2018MNRAS.475..624N, 2018MNRAS.477.1206N, 2018MNRAS.475..676S, 2018MNRAS.475..648P, 2019ComAC...6....2N}. Specifically, they found that the quenched fraction of satellite galaxies with \(M_* > 10^{9.5} M_{\odot}\) is significantly lower than our measurements, whereas at \(M_* < 10^{9.5} M_{\odot}\), it rises sharply and exceeds our observed values. While \citetalias{2024ApJ...969..129Z} could only compare the total quenched fraction of satellites, our new results allow us to disentangle the effects of mass and environment separately.

We reproduce all results in Figure \ref{central_fq} using \texttt{TNG300-1} by counting red and blue galaxies in each component. To ensure consistency with observations, we use the catalog containing synthetic stellar photometry (i.e., colors) provided by \citep{2018MNRAS.475..624N} and apply the same color cut as in the observations. Stellar mass is defined as the sum of all stellar particles within twice the stellar half-mass radius. The results from \texttt{TNG300-1} are also shown in Figure \ref{central_fq}. Within the studied stellar mass range (\([10^{9.5}M_{\odot},10^{11.0}M_{\odot}]\)), the mass-quenched fraction in \texttt{TNG300-1} is progressively lower than our findings as stellar mass decreases. This suggests potential deficiencies in the modeling of internal quenching processes, such as AGN and supernova feedback. Meanwhile, the environmental quenched fraction in \texttt{TNG300-1} aligns remarkably well with our results. However, this agreement may be coincidental and does not necessarily indicate that \texttt{TNG300-1} correctly models environmental quenching, as the relevant quantity is QFE (Equation \ref{eq:QFE}) rather than the absolute fraction $f_{\rm{q}}^{\rm{env}}$. Since $\bar{f}_{\rm{q}}^{\rm{cen}}$ in \texttt{TNG300-1} is lower than in observations, QFE should be higher, implying that environmental quenching in \texttt{TNG300-1} is actually more efficient than observed. As our study does not extend to \(M_{*,{\rm sat}}<10^{9.5}M_{\odot}\) to avoid extrapolation risks, determining whether the upturn in the satellite quenched fraction at lower masses in \texttt{TNG300-1} results from deficiencies in environmental quenching, mass quenching, or both will require more precise measurements.      

\section{Conclusion} \label{sec:Coclusion}
In this paper, building on the \(\bar{n}_2w_{\rm{p}}\) measurements from \citetalias{2024ApJ...969..129Z}, we develop a method to disentangle mass and environmental quenching. The key components of this method include reconstructing the 3D quenched fraction distribution, \(f^{\rm{sat}}_{\mathrm{q}}(r;M_{*,\rm{cen}},M_{*,\rm{sat}})\), for satellite galaxies and determining the average quenched fraction, \(\bar{f}_{\rm{q}}^{\rm{cen}}(M_{*})\), for central galaxies, which represents the mass-quenched fraction. The latter is achieved by combining the \(f^{\rm{sat}}_{\mathrm{q}}(r;M_{*,\rm{cen}},M_{*,\rm{sat}})\) model with the galaxy-halo connection in N-body simulations. Using \(f^{\rm{sat}}_{\mathrm{q}}(r;M_{*,\rm{cen}},M_{*,\rm{sat}})\), \(\bar{f}_{\rm{q}}^{\rm{cen}}(M_{*})\), and the galaxy-halo connection, we assign a quenched probability to each galaxy in the simulation, enabling a comprehensive study of galaxy quenching. Our principal findings are summarized as follows:

\begin{itemize}
    \item Mass quenching dominates the entire stellar mass range \([10^{9.5}, 10^{11.0}]M_{\odot}\) studied. The mass-quenched fraction increases from 0.3 to 0.87 across the stellar mass range \([10^{9.5}, 10^{10.0}]M_{\odot}\) to \([10^{10.5}, 10^{11.0}]M_{\odot}\), while the environmental quenched fraction decreases from 0.17 to 0.03.
    \item  More massive host halos are more effective at quenching their satellite galaxies, while satellite stellar mass has minimal influence on environmental quenching, as indicated by the QFE.
    \item Within the studied stellar mass range (\([10^{9.5}M_{\odot},10^{11.0}M_{\odot}]\)), \texttt{TNG300-1} exhibits lower mass quenching efficiency but higher environmental quenching compared to observations.
\end{itemize}

In this study, we only have measurements to validate our model in environments with \(M_{\rm{vir}} > 10^{12.0} h^{-1} M_{\odot}\). However, with our method, we can extend the exploration of galaxy quenching to lower stellar and host halo masses using next-generation galaxy surveys.

\section*{Acknowledgment}
The work is supported by NSFC (12133006), by National Key R\&D Program of China (2023YFA1607800, 2023YFA1607801), and by 111 project No. B20019. This work made use of the Gravity Supercomputer at the Department of Astronomy, Shanghai Jiao Tong University. K.X. is supported by the funding from the Center for Particle Cosmology at U Penn. Y.P.J. gratefully acknowledge the support of the Key Laboratory for Particle Physics, Astrophysics and Cosmology, Ministry of Education. We acknowledge the science research grants from the China Manned Space Project with NO. CMS-CSST-2021-A03. 

This publication has made use of data products from the Sloan Digital Sky Survey (SDSS). Funding for SDSS and SDSS-II has been provided by the Alfred P. Sloan Foundation, the Participating Institutions, the National Science Foundation, the U.S. Department of Energy, the National Aeronautics and Space Administration, the Japanese Monbukagakusho, the Max Planck Society, and the Higher Education Funding Council for England.

\appendix
\restartappendixnumbering
\section{Fitting Models to PAC Measurements}\label{sec:fitting}
\begin{figure*}
\begin{center}
\includegraphics[width=1\textwidth]{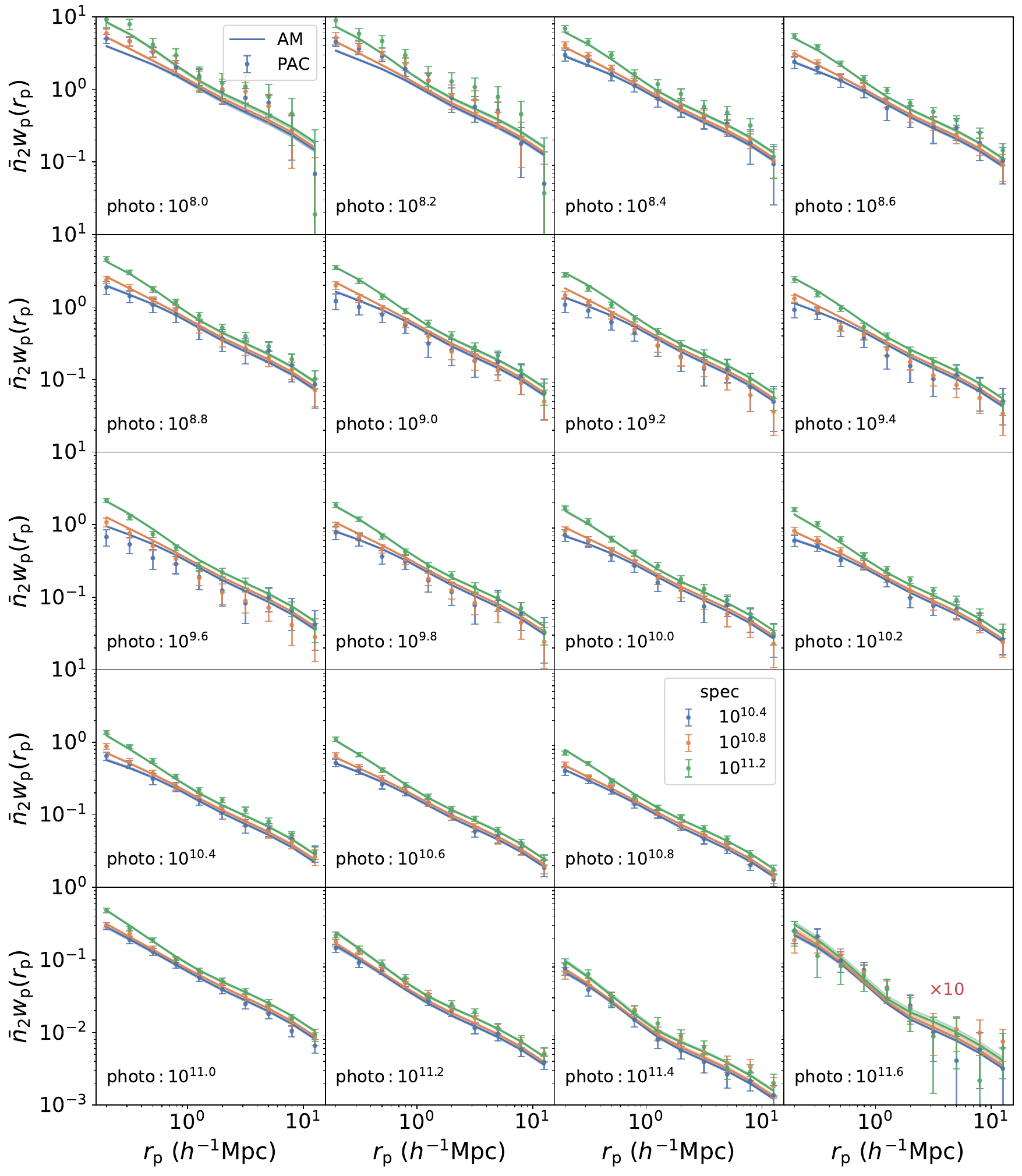}
\end{center}
\caption{PAC measurements from \citetalias{2023ApJ...944..200X} and best-fit results from the SHAM model in \texttt{Jiutian} simulation using the SHMR shown in Figure \ref{SHMR}. Dots with error bars represent the measurements, while solid lines with shaded regions indicate the best-fit results and corresponding \(1\sigma\) uncertainties from SHAM modeling. For clarity, we display only three \({\rm{pop}}_1\) mass bins (\(10^{10.4}, 10^{10.8}, 10^{11.2}M_{\odot}\)) out of the total five. Additionally, results for the final \({\rm{pop}}_2\) bin (\(10^{11.6}M_{\odot}\)) are scaled by a factor of 10 for better visualization.}
\label{fig:PAC_fit}
\end{figure*}

In Figure \ref{fig:PAC_fit}, we show the SHAM model fits, based on the stellar mass–halo mass relation (SHMR) from Figure \ref{SHMR} in the \texttt{Jiutian} simulation, to the PAC measurements. The fits to the 95 cross-correlations are generally excellent, with a reduced chi-square value of \(\chi^2/{\rm{d.o.f.}} \sim 0.8\).

\bibliography{sample631}{}
\bibliographystyle{aasjournal}



\end{document}